\documentclass{article}
\usepackage{amsmath,amssymb,amscd,latexsym,amsthm}
\newtheorem{define}{Definition}
\newtheorem{theorem}{Theorem}
\theoremstyle{conjecture}
\newtheorem*{conjecture}{Conjecture}
\def\Slash#1{#1\!\!\!\!/}
\def\Dirac{\Slash D}
\author{Peter Woit \\
Department of Mathematics, Columbia University\\
woit@math.columbia.edu}
\title{Quantum Field Theory and Representation Theory: A Sketch}
\begin{document}
\maketitle
\tableofcontents
\section{Introduction}

Ever since the early days of theory there has been a close link
between representation theory and quantum mechanics. The Hilbert
space of quantum mechanics is a (projective) unitary representation of the
symmetries of the classical mechanical system being quantized. The
fundamental observables of quantum mechanics correspond to the
infinitesimal generators of these symmetries (energy corresponds
to time translations, momentum to space translations, angular
momentum to rotations, charge to phase changes). The relation
between quantum mechanics and representation theory has been
formalized as the subject of \lq\lq geometric quantization" which
ideally associates to a classical mechanical phase space (a
symplectic manifold M) a complex vector space V in a functorial
manner. This functor takes at least some subgroup $G$ of the
symplectomorphisms (canonical transformations) of M to unitary
transformations of V, making V a unitary $G$-representation.

The theory of geometric quantization has never been very popular among physicists for at least
two reasons (in addition to the fact that the mathematical apparatus required is rather
extensive and mostly unfamiliar to physicists). The first is that it seems to have very little
to say about quantum field theory.  The quantum field theory of the standard model of particle
physics is built upon the geometrical concepts of gauge fields and the Dirac operator on
spinors, concepts which have no obvious relation to those used in geometric quantization.

The second problem is that the most well-developed formalism for doing calculations within
the standard model is the path integral formalism.  While no rigorous version of this exists,
all evidence is that consistent calculations can be performed using this formalism, at least
within perturbation theory or outside of perturbation theory with a lattice cut-off. Even in the
simple case of quantum mechanics the relationship between the path integral quantization and
geometric quantization has been quite unclear making it impossible to see how the ideas of
geometric quantization can be useful in the much more complex situation of standard model
quantum field theory.

Taking the path integral as fundamental, in its sketchiest form the problem of understanding
the standard model quantum field theory comes down to that of making sense
of ratios of expressions such as

$$\int [dA]  (\int [d\Psi] O(A,\Psi) e^{\int_M
\Psi\Dirac_A\Psi})e^{\int_M -\frac{1}{g^2} ||F_A||^2}$$

Here $\int [dA]$ is supposed to be an integral over the space of connections on
a principal bundle over the manifold $M=\mathbf R^4$. The variables  $\Psi$ are
sections of some product of vector bundles, one factor of which is the spinor bundle of M. $\int [d\Psi]$ is supposed to
be the linear functional on an infinite dimensional exterior algebra generated by variables $\Psi$ given
by taking the coefficient of the top dimensional power of the $\Psi$'s, $\int_M ||F_A||^2$
is the norm-squared of the curvature of the connection $A$ and $\Dirac_A$ is the
Dirac operator formed using the covariant derivative determined by $A$. 
$O(A,\Psi)$
is some functional of the connection and the $\Psi$ variables. One only really expects
to make sense of this expression for certain classes of $O(A,\Psi)$.  To make such path
integrals well-defined one needs to choose a distance scale called a \lq\lq cut-off"
and suppress integration over
variables that vary on scales smaller than the cut-off.  The parameter $g$ depends on the
cut-off and for an asymptotically free theory must be taken to zero in a specified manner
as the cut-off goes to zero.

To a mathematician, such path integrals immediately raise a host of questions.
\begin{itemize}
\item
Why is the space of connections
of a principal bundle appearing and why is one trying to integrate over it?
\item
What is the
significance of the Dirac operator acting on sections of the spinor bundle? Does this
have anything to do with the same structures that appear in index theory?
\item
Why is one considering
$$e^{\int_M\Psi\Dirac_A\Psi}$$
and extracting
the coefficient of the top dimensional power of the infinite dimensional exterior algebra over the spinor fields?
\end{itemize}

The present work is motivated by the desire to try and get some answers to these
questions by investigating what sort of formal mathematical problem
such path integral expressions could represent the solution to.  A hint is provided by the deep
relationship between quantum theory and representation theory exposed by the theory of geometric
quantization.  In brief, we are looking for a representation-theoretical interpretation of the
kind of quantum field theory that appears in the standard model. The groups involved will be
ones whose representation theory is not at all mathematically understood at the present time so
unfortunately known mathematics cannot be used to say much about these quantum field theories.
On the other hand, the wealth of knowledge that physicists have accumulated about these theories
may be useful to mathematicians in trying to understand something about the representation
theory of certain important infinite dimensional groups.

The main goal of this paper is to explain and provide a sketch of evidence for the following
conjecture:
\begin{conjecture}
The quantum field theory of the standard model may be understood purely in terms
of the representation theory of the automorphism group of some geometric
structure.
\end{conjecture}

The reader is to be warned that the present version of this document suffers from sloppiness
on several levels.  If factors of $2$, $\pi$ and $i$ seem to be wrong, they probably are.
Many of the mathematical statements are made with a blatant disregard for mathematical precision.
Sometimes this is done out of ignorance, sometimes out of a  desire to simply get to
the heart of the matter at hand.  The goal has been to strike a balance such that physicists may
have a fighting chance of reading this while mathematicians may not find the level of imprecision 
and simplification too hard to tolerate.

\section{Quantizing G/T: The Representation Theory of Compact Lie Groups}

The standard description in physics textbooks of how to quantize a
Hamiltonian classical mechanical system instructs one to first
choose \lq\lq canonical coordinates", (i.e. coordinates $q_i, p_i$ for $i=1,\dots,n$) on phase space
${\mathbf R}^{2n}$ satisfying
$$\{q_i, q_j\}=0, \{p_i,p_j\}=0, \{q_i,p_j\}=\delta_{ij}$$
where $\{\cdot,\cdot\}$ is the Poisson bracket. Equivalently, one is choosing a
symplectic structure
$$\omega=\sum_{i=1}^n dq_i\wedge dp_i$$
on ${\mathbf R}^{2n}$.

The corresponding quantum system then is defined in terms of
a Hilbert space $\mathcal H$ and operators corresponding to the
canonical coordinates satisfying
$$[\hat{q_i}, \hat{q_j}]=0, [\hat{p_i},\hat{p_j}]=0, [\hat{q_i},\hat{p_j}]=i\hbar\delta_{ij}$$
One can construct various explicit Hilbert spaces $\mathcal H$ and sets of operators
on $\mathcal H$ satisfying these conditions, for instance by taking
$\mathcal H$ to be a set of functions on the coordinates $q_i$, $\hat{q_i}$ to be multiplication
by $q_i$, and $\hat{p_i}$ to be the operator $-i\hbar\frac{\partial}{\partial q_i}$.

A more abstract way of looking at this is to note that the
$\hat{q_i},\hat{p_i}$ satisfy the defining relations of the Lie
algebra of the Heisenberg group and $\mathcal H$ should be a
(projective) unitary representation of this group. The Stone-von Neumann
theorem tells one that up to equivalence this representation is
unique and implies that linear symplectic transformations act
(projectively) on $\mathcal H$. This projective representation is
a true representation of the metaplectic group, a $\mathbf Z_2$
extension of the symplectic group.

While this procedure works fine for phase spaces that can be globally identified with $\mathbf
R^{2n}$, it fails immediately in other simple cases.  For instance, the phase space of a fixed
spinning particle of spin $\frac{k}{2}$ in $\mathbf R^3$ can be taken to be the sphere $S^2$
with area $k$, but there is no way to globally choose canonical
coordinates in this case.  The corresponding quantum theory should have ${\mathcal H} = {\mathbf
C}^{k+1}$ with angular momentum operators $L_i, i=1,2,3$ satisfying the commutation relations of
the Lie algebra of the group Spin(3).

While the standard physics quantization procedure fails in this
simple case, it is an example of a well-known phenomenon in the
representation theory of Lie groups.  In this section we will
review the representation theory of a compact connected Lie group
$G$ from a point of view which makes clear the relationship
between the representation and the corresponding phase space.
Quite a lot of modern mathematics goes into this story and
certainly is not needed for the problem at hand, but has been
developed to deal with more general cases such as that of
non-compact groups.

\subsection{The Borel-Weil Theorem}

Let $G$ be a compact, connected Lie group, it will carry a left and right invariant Haar measure.  A
copy ($G_L$) of $G$ acts from the left on the Hilbert space $L^2(G)$ as the left regular representation.
This representation is infinite dimensional and reducible. There is a commuting action on the right by
another copy ($G_R$) of $G$. In the case of $G=U(1)$ the decomposition of the regular representation
into irreducible representations is given by Fourier analysis. The irreducible representations are
one-dimensional and labelled by an integer. More generally:

\begin{theorem}[Peter-Weyl]
There is a
Hilbert space direct sum decomposition
$$L^2(G)=\sum_{i\in \hat G}V_i\otimes V_i^*$$
where $\hat G$ is the set of irreducible representations
of $G$.
\end{theorem}

Here $V_i$ is an irreducible representation of $G$ on a complex
vector space of dimension $d_i$ and $V_i^*$ is the dual
representation. $G_R$ acts on the $V_i^*$ factor, $G_L$ acts on
the $V_i$ factor.  Thus the irreducible $V_i$ occurs in either the
right or left regular representation of $G$ on $L^2(G)$ with
multiplicity $d_i$.

One can understand this decomposition as a decomposition of functions on $G$ into matrix
elements of irreducible representations. To the element $$(|\alpha>,<\beta|)\in V_i\times
V_i^*$$
 is associated the function on $G$
given by $$f(g)=<\beta |\pi_i(g^{-1}) |\alpha>$$ where $\pi_i(g)$
is the representation of the group element $g$ on the vector space
$V_i$.  Under this association the map
$$f(g)\rightarrow f(g_L^{-1}g)$$ corresponds to
$$|\alpha>\rightarrow \pi_i(g_L)|\alpha>$$
and
$$f(g)\rightarrow f(gg_R)$$ corresponds to
$$<\beta|\rightarrow <{\overline \pi}_i(g)\beta|$$
where ${\overline \pi}_i(g)$ is the contragredient representation
to $\pi_i(g)$.

The problem still remains of identifying the irreducible
representations of $G$ and computing their dimensions $d_i$. This
can be done by relating irreducible representations of $G$ to
representations of the maximal torus $T$ of $G$, whose
representations are given by Fourier analysis.  This can be done
with the Cartan-Weyl theory of the highest weight. Detailed
expositions of the theory can be found in \cite{Adams,Br-tD}.

Different choices of maximal tori $T$ are related by conjugation by an element $g\in G$
$$T\rightarrow gTg^{-1}$$
The subgroup of $G$ that leaves $T$ invariant under this conjugation is called the
normalizer $N(T)$ and:
\begin{define}
The Weyl group $W(G,T)$ is the group $N(T)/T$ of non-trivial automorphisms of $T$ that
come from conjugations in $G$.
\end{define}

The character $\chi_V$ of a representation is a conjugation
invariant function on $G$ which can be computed in a matrix
representation $\pi_V$ as a trace
$$\chi_V(g) = tr_{\pi_V}(g)$$
As a function on $T$, $\chi_V$ will always be invariant under the action of $W(G,T)$ on $T$.

In order to find an explicit decomposition of $L^2(G)$ into irreducible representations
we will begin by decomposing
$L^2(G)$ into pieces that transform under the action of $T$ from the right according to the various
weights of $T$. Picking a weight $\lambda$ of $T$, consider the subspace of
$L^2(G)$ that satisfies
$$f(gt)=t^{-1}\cdot f(g)=\chi_\lambda^{-1}f(g)$$

An equivalent definition of this
space is as a space of sections $\Gamma(L_\lambda)$ of the line bundle
\begin {equation*}
\begin{CD}
{\mathbf C} @>>>G\times_T{\mathbf C}=L_\lambda \\
@. @VVV\\
{} @. G/T
\end{CD}
\end{equation*}
over the quotient manifold $G/T$ (sometimes known as a flag
manifold since in the case $G=U(n)$ it is the space of flags in
$\mathbf C^n$).

$L_\lambda$ has a curvature 2-form $\omega_\lambda$ which is
invariant under the left $G$ action on $G/T$.  This is a
symplectic structure, so $G/T$ can be thought of as the phase
space for a mechanical system.  In the simple case $G=SU(2)$,
$T=U(1)$, the $\lambda$ are labelled by the integers and
$$L_{\lambda}=SU(2)\times_{U(1)}{\mathbf C}$$ is
the $\lambda$'th power of the tautological line bundle over ${\mathbf CP_1}$.

The left action of $G$ on $L^2(G)$ leaves $\Gamma(L_\lambda)$
invariant and $\Gamma(L_\lambda)$ is an infinite dimensional
reducible representation.  This representation is sometimes
thought of as the \lq\lq pre-quantization" of the symplectic manifold
$G/T$ with symplectic form $\omega_\lambda$.
In terms of the decomposition into
matrix elements of the Peter-Weyl theorem, $\Gamma(L_\lambda)$ is
the subspace of matrix elements that are sums over all irreducible
representations of terms of the form
$$<\beta |\pi_i^{-1}(gt) |\alpha>=<\beta |\pi_i^{-1}(t)\pi_i^{-1}(g) |\alpha>=
<\beta |\chi_\lambda^{-1}(t)\pi_i^{-1}(g) |\alpha>$$
i.e. the matrix elements such that $<\beta|$ is in the subspace
$V_{i,\lambda}^*$ of $V_i^*$ that transforms with character
$\chi_\lambda$ under $T$. $V_{i,\lambda}^*$ can equivalently be defined as
$$(V_i^*\otimes \mathbf C_\lambda)^T$$

We have shown that $\Gamma(L_\lambda)$ decomposes under the left $G$
action into a direct sum of irreducibles $V_i$ with the
multiplicity given by the dimension of $V_{i,\lambda}^*$.
$$\Gamma(L_\lambda)=\sum_{i\in \hat G}V_i\otimes (V_i^*\otimes \mathbf C_\lambda)^T$$

The true
quantization should be an irreducible representation of $G$ and
this will require using further structure.
To pick out a single irreducible we need to consider the Lie
algebra $\mathfrak g$ and how it transforms as a real
representation under the adjoint action of $T$. The Lie subalgebra
$\mathfrak t$ of $\mathfrak g$ corresponding to $T$ transforms
trivially under this action. $T$ acts without invariant subspace
on the quotient ${\mathfrak g}/{\mathfrak t}$ so it breaks up into
two-dimensional subspaces.  Thus ${\mathfrak g}/{\mathfrak t}$ is
an even-dimensional real vector space.  While ${\mathfrak
g}/{\mathfrak t}$ is not a Lie algebra, a choice of complex
structure on this space gives a decomposition of the
complexification as

$$ {\mathfrak g}/{\mathfrak t}\otimes{\mathbf C}={\mathfrak n}_+
\oplus {\mathfrak n}_-$$

Here
$${\mathfrak n}_+=\sum_{\alpha\in \Phi} {\mathfrak g}^\alpha$$ where $\Phi$ is a set
labelling the so-called positive roots and ${\mathfrak n}_-$ is
its complex conjugate. The ${\mathfrak g}^\alpha$ are the various
one-dimensional complex root spaces on which $T$ acts with weight
$\alpha$.

Different choices of the set $\Phi$ corresponding to different
invariant complex structures are in one-to-one correspondence with
elements of the Weyl group $W(G,T)$.

Globally over $G/T$ one can form
\begin{equation*}
\begin{CD}
{\mathfrak g}/{\mathfrak t}\otimes{\mathbf C}@>>>G\times_T ({\mathfrak g}/{\mathfrak t}\otimes{\mathbf C})\\
@. @VVV\\
{} @. G/T
\end{CD}
\end{equation*}
which is the complexified tangent bundle of $G/T$.  The
decomposition into positive and negative roots gives an integrable
complex structure since
$$[{\mathfrak n}_+,{\mathfrak n}_+]\subset {\mathfrak n}_+$$
and
$$[{\mathfrak n}_-,{\mathfrak n}_-]\subset {\mathfrak n}_-$$

So for each element of the Weyl group $G/T$ is a complex manifold,
but in an inequivalent way.  $L_\lambda$ is a holomorphic line bundle
and one can show that the holomorphic sections of $L_\lambda$ are precisely
those sections of $\Gamma(L_\lambda)$ that are invariant under infinitesimal
right translations generated by elements of $\mathfrak n_-$, i.e.

$$\Gamma^{hol}(L_\lambda)=\{f\in \Gamma(L_\lambda): r(X)f=0\ \forall X\in \mathfrak n_-\}$$
or
$$\Gamma^{hol}(L_\lambda)=\sum_{i\in \hat G}V_i\otimes \{v\in(V_i^*\otimes \mathbf C_\lambda)^T
:{\mathfrak n}_- v=0\}$$

The condition ${\mathfrak n}_- v=0$ picks out the lowest weight space in $V_i^*$, so the
sum on the right is zero unless the lowest weight space of $V_i^*$ has weight $-\lambda$ or
equivalently the highest weight of $V_i$ is $\lambda$.  The classification theory of representations
by their highest weight implies that for each $\lambda$ in the dominant Weyl chamber there is
precisely one irreducible representation with highest weight $\lambda$.  For $\lambda$ not in the
highest weight chamber there are no such representations.  So we have

\begin{theorem}[Borel-Weil]
For $\lambda$ dominant $\Gamma^{hol}(L_\lambda)=H^0(G/T,{\mathcal
O}(L_\lambda))$ is the irreducible $G$ representation $V_\lambda$
of highest weight $\lambda$, and for $\lambda$ not dominant this
space is zero.  This gives all finite dimensional irreducible
representations of $G$.
\end{theorem}

By demanding that functions on $G$ transform as $\lambda$ under
the right $T$ action and using the rest of the right $G$ action to
impose invariance under infinitesimal right ${\mathfrak n}_-$
translations, we have constructed a projection operator on
functions on $G$ that gives zero unless $\lambda$ is dominant.
When $\lambda$ is dominant, this projection picks out a space of
functions on $G$ that transforms under the left $G$ action as the
irreducible representation with highest weight $\lambda$.

\subsection{Lie Algebra Cohomology and the Borel-Weil-Bott-Kostant Theorem}

The Borel-Weil theorem gives a construction of the irreducible
representations of $G$ as $\Gamma(L_\lambda)^{\mathfrak{n}_-}$,
the ${\mathfrak{n}_-}$ invariant part of $\Gamma(L_\lambda)$.  The
definition of ${\mathfrak{n}_-}$ depends upon a choice of
invariant complex structure. A non-trivial element $w$ of the Weyl
group $W(G,T)$ gives a different choice of $\mathfrak{n}_-$ (call
this one $\mathfrak{n}_-^w$) and now
$\Gamma(L_\lambda)^{\mathfrak{n}_-^w} =0$ for a dominant weight
$\lambda$.  Work of Bott \cite{Bott1} and later
Kostant \cite{Kostant1} shows that the representation now occurs in
a higher cohomology group.  This is one motivation for the
introduction of homological algebra techniques into the study of
the representation theory of $G$.

To any finite dimensional representation $V$ of $G$, one can associate a more algebraic
object, a module over the enveloping algebra $U(\mathfrak g)$, which we will also call $V$.
If elements of the Lie algebra $\mathfrak g$ are thought of as the left-invariant vector
fields on $G$, then elements of $U(\mathfrak g)$ will be the left-invariant partial
differential operators on $G$.  In homological algebra a fundamental idea is to
replace the study of a $U(\mathfrak g)$ module $V$ with a resolution of $V$, a complex
of $U(\mathfrak g)$ modules where each term has simpler properties, for instance that
of being a free $U(\mathfrak g)$ module.

\begin{define}
A resolution of $V$ as a $U(\mathfrak g)$ module is an exact
sequence
$$0\leftarrow V \leftarrow M_0\leftarrow M_1 \leftarrow \cdots \leftarrow M_n\leftarrow 0$$
of $U(\mathfrak g)$ modules and $U(\mathfrak g)$-linear maps.  It is
a free (resp. projective, injective) resolution if the $M_i$ are free (resp. projective,
injective) $U(\mathfrak g)$ modules.
\end{define}

Note that deleting $V$ from this complex give a complex whose homology is simply
$V$ in degree zero.  One is essentially replacing the study of $V$ with the study
of a \lq\lq quasi-isomorphic" complex whose homology is $V$ in degree zero.  Now the trivial $U(\mathfrak g)$
module $V=\mathbf C$
becomes interesting, it has a free resolution known as the standard or Koszul
resolution:
\begin{define}
The Koszul resolution is the exact sequence of $U(\mathfrak g)$ modules
$$0\longleftarrow\mathbf C\stackrel{\epsilon}\longleftarrow Y_0\stackrel{\partial_0}\longleftarrow Y_1
\stackrel{\partial_1}\longleftarrow\cdots\longleftarrow Y_{n-1}\stackrel{\partial_{n-1}}\longleftarrow Y_n
\longleftarrow 0$$
where
$$Y_m=U(\mathfrak g)\otimes_{\mathbf C}\Lambda^m(\mathfrak g)$$
$$\epsilon(u)= {\mathrm {constant\ term\ of\ }}u\in U(\mathfrak g)$$
and
$$\partial_{m-1}(u\otimes X_1\wedge X_2\wedge\cdots\wedge X_m)=\sum_{i=1}^{m}
(-1)^{i+1}(uX_i\otimes X_1\wedge\cdots\wedge\hat X_i\wedge\cdots\wedge X_m)$$
$$+\sum_{k<l}(-1)^{k+l}(u\otimes[X_k,X_l]\wedge X_1\wedge\cdots
\wedge \hat X_k\wedge\cdots \wedge \hat X_l\wedge\cdots \wedge X_m)$$
(Here $X_i\in \mathfrak g$ and $\hat X_i$ means delete $X_i$ from the wedge product)

\end{define}

Applying the functor $\cdot\rightarrow Hom_{\mathbf C}(\cdot, V)$ to
the Koszul complex we get a new exact sequence (the functor is "exact")

$$0\longrightarrow V\longrightarrow Hom_{\mathbf C}(Y_0,V)\longrightarrow\cdots
\longrightarrow Hom_{\mathbf C}(Y_n,V)\longrightarrow 0$$
which is a resolution of $V$.  The $\mathfrak g$-invariant part of a $U(\mathfrak g)$ module
can be picked out by the \lq\lq invariants functor"
$$\cdot\longrightarrow Hom_{U(\mathfrak g)}(\mathbf C,\cdot)= (\cdot)^{\mathfrak g}$$
which takes $U(\mathfrak g)$ modules to vector spaces over $\mathbf C$.  Applying this
functor to the above resolution of $V$ gives the complex
$$0\longrightarrow V^{\mathfrak g} \longrightarrow C^0(\mathfrak g,V)\longrightarrow
C^1(\mathfrak g,V)\longrightarrow\cdots\longrightarrow C^n(\mathfrak g,V)\longrightarrow 0$$
where
\begin{eqnarray*}
C^m(\mathfrak g,V)&=&Hom_{U(\mathfrak g)}(\mathbf C,Hom_{\mathbf C}(Y_m,V))\\
&=&(Hom_{\mathbf C}(Y_m,V))^{\mathfrak g}\\
&=&Hom_{U(\mathfrak g)}(Y_m,V)\\
&=&Hom_{\mathbf C}(\Lambda^m(\mathfrak g),V)
\end{eqnarray*}

The invariants functor is now no longer exact and Lie algebra
cohomology is defined as its "derived functor" given by the
homology of this sequence (dropping the $ V^{\mathfrak g}$ term).
More explicitly

\begin{define}
The Lie algebra cohomology groups of $\mathfrak g$ with coefficients in the $U(\mathfrak g)$ module $V$
are the groups
$$H^m(\mathfrak g, V)=\frac{\mathrm{Ker\ }d_m}{\mathrm{Im\ }d_{m-1}}$$
constructed from the complex
$$0\longrightarrow V\stackrel{d_0}\longrightarrow \Lambda^1(\mathfrak g)\otimes_{\mathbf C}V
\stackrel{d_1}\longrightarrow\cdots\stackrel{d_{n-1}}\longrightarrow\Lambda^n(\mathfrak g)\otimes_{\mathbf C}V
\longrightarrow 0$$
where the operator $d_n : C^n(\mathfrak g,V)\rightarrow C^{n+1}(\mathfrak g,V)$ is given by
$$d_n\omega(X_1\wedge\cdots\wedge X_{n+1})=\sum_{l=1}^{n+1}(-1)^{l+1}X_l(\omega(X_1\wedge\cdots\wedge\hat X_l
\wedge\cdots\wedge X_{n+1}))$$
$$+\sum_{r<s}(-1)^{r+s}\omega([X_r,X_s]\wedge X_1\wedge\cdots\wedge\hat X_r\wedge\cdots\wedge\hat X_s
\wedge\cdots\wedge X_{n+1})$$
They satisfy
$$H^0(\mathfrak g)=V^{\mathfrak g}$$
\end{define}

For the case $V=\mathbf C$ and $G$ a compact Lie group, the $H^m({\mathfrak g},\mathbf C)$ correspond
with the de Rham cohomology groups of the topological space $G$, this is not true in general
for non-compact groups.

More generally one can consider the functor
$$\cdot\longrightarrow Hom_{U(\mathfrak g)}(W,\cdot)$$
for some irreducible representation $W$ of $G$. This functor takes $U(\mathfrak g)$
modules to a vector space of dimension given by the multiplicity of the irreducible
$W$ in $V$.  The corresponding derived functor is
$Ext^*_{U(\mathfrak g)}(W,V)$ and is equal to
$H^*(\mathfrak g, Hom_{\mathbf C}(W,V))$.

In the previous section we considered $\Gamma (L_\lambda)$ as a $U(\mathfrak n_-)$
module using the right $\mathfrak n_-$ action on functions on $G$.  The
$\mathfrak n_-$-invariant part of this module was non-zero and an irreducible
representation $V_\lambda$ under the left $G$-action.  For a non-dominant
weight $\lambda$ the $\mathfrak n_-$-invariant part was zero, but it turns out
we can get something non-zero by replacing the  $\Gamma (L_\lambda)$ by its
resolution as a $U(\mathfrak n_-)$ module.  Kostant computed the Lie algebra
cohomology in this situation, finding \cite{Kostant1}
\begin{theorem}[Kostant]

As a $T$ representation
$$H^*(\mathfrak n_-,V_\lambda)=\sum_{w\in W(G,T)} \mathbf C_{w(\lambda+\delta)-\delta}$$
i.e. the cohomology space is a sum of one-dimensional $T$ representations, one for each
element of the Weyl group, transforming under $T$ with weight $w(\lambda+\delta)-\delta$ where
$\delta$ is half the sum of the positive roots.  Each element $w$ is characterized by
an integer $l(w)$, its length, and $ C_{w(\lambda+\delta)-\delta}$ occurs in degree
$l(w)$ of the cohomology space.
\end{theorem}

Replacing  $\Gamma (L_\lambda)$ by its resolution as a $U(\mathfrak n_-)$ module gives the
$\bar\partial$ complex $\Omega^{0,*}(G/T,{\mathcal O}(L_\lambda))$ that computes not just
$\Gamma^{hol}(L_\lambda)=H^0(G/T,{\mathcal
O}(L_\lambda))$ but all the cohomology groups $H^*(G/T,{\mathcal O}(L_\lambda))$ with the
result

\begin{theorem}[Borel-Weil-Bott-Kostant]

$$H^*(G/T, {\mathcal O}(L_\lambda))=H^{0,*}(\Omega^{\cdot,\cdot}(G/T,{\mathcal O}(L_\lambda)))
=\sum_{i\in \hat G}V_i\otimes (H^*({\mathfrak n_-},V_i^*)\otimes C_\lambda)^T$$
For a given weight $\lambda$, this is non-zero only in degree $l(w)$ where $w(\lambda+\delta)$
is dominant and gives an irreducible representation of $G$ of highest weight $w(\lambda +\delta)
-\delta$.
\end{theorem}

Knowledge of the T-representation structure of the complex that computes $H^*(\mathfrak
n_-,V_\lambda)$ allows one to quickly derive the Weyl character formula.  The character
of $\sum_{i=0}^n(-1)^iC^i(\mathfrak n_-,V_\lambda)$ is the product of the character of
$V(\lambda)$ and the character of $\sum_{i=0}^n(-1)^iC^i(\mathfrak n_-,\mathbf C)$ so

$$\chi_{V_\lambda}=\frac{\sum_{i=0}^n(-1)^i \chi_{C^i(\mathfrak n_-,V_\lambda)}}
{\sum_{i=0}^n(-1)^i \chi_{C^i(\mathfrak n_-,\mathbf C)}}$$

By the Euler-Poincar\'e principle the alternating sum of the characters of terms
in a complex is equal to the alternating sum of the cohomology groups, the Euler
characteristic, so

$$\chi_{V_\lambda}=\frac{\sum_{i=0}^n(-1)^i \chi_{H^i(\mathfrak n_-,V_\lambda)}}
{\sum_{i=0}^n(-1)^i \chi_{H^i(\mathfrak n_-,\mathbf C)}}=\frac{\chi({\mathfrak n}_-,V_\lambda)}
{\chi({\mathfrak n}_-,\mathbf C)}$$

Where on the right $\chi$ denotes the Euler characteristic, as a $T$-representation.  Using
the explicit determination of the Lie algebra cohomology groups discussed above, one gets
the standard Weyl character formula.

The introduction of homological methods in this section has allowed us to construct irreducible
representations in a way that is more independent of the choice of complex structure on
$G/T$.  Under change of complex structure, the same representation will occur, just in
a different cohomological degree.  These methods also can be generalized to the case of
discrete series representations of non-compact groups, where there may be no complex structure for
which the representations
occurs in degree zero as holomorphic sections.  The general idea of constructing representations
on cohomology groups is also important in other mathematical contexts, for instance in number
theory where representations of the Galois group $Gal({\mathbf {\bar Q}}/{\mathbf Q})$ are constructed
on $l$-adic cohomology groups of varieties over ${\mathbf {\bar Q}}$.

\subsection{The Clifford Algebra and Spinors: A Digression}

The Lie algebra cohomology construction of the irreducible
representations of $G$ requires choosing an invariant complex
structure on $\mathfrak g/\mathfrak t$, although ultimately
different complex structures give the same representation. It
turns out that by using spinors instead of the exterior powers
that appear in the Koszul resolution, one can construct
representations by a method that is completely independent of the
complex structure and that furthermore explains the somewhat
mysterious appearance of the weight $\delta$. For more on this 
geometric approach to spinors see \cite{Chevalley} and
\cite{PS}.

\subsubsection{Clifford Algebras}

Given a real $2n$-dimensional vector space $V$ with an inner product $(\cdot,\cdot)$, the Clifford
algebra $C(V)$ is the algebra generated by the elements of $V$ with multiplication
satisfying
$$v_1v_2+v_2v_1=2(v_1,v_2)$$
In particular $v^2=(v,v)=||v||^2$.  $C(V)$ has a $\mathbf
Z_2$-grading since any element can be constructed as a product of
either an even or odd number of generators. For the case of zero
inner product $C(V)=\Lambda^*(V)$, the exterior algebra, which has
an integer grading.  In some sense the Clifford algebra is more
basic than the exterior algebra, since given $C(V)$ one can
recover $\Lambda^*(V)$ as the associated graded algebra to the
natural filtration of $C(V)$ by the minimal number of generating
vectors one must multiply to get a given element.

The exterior algebra $\Lambda^*(V)$ is a module over the Clifford algebra, taking
Clifford multiplication by a vector $v$
to act on $\Lambda^*(V)$ as
$$\cdot \rightarrow v\wedge\cdot + i_v(\cdot)$$
Here $i_v(\cdot)$ is interior multiplication by $v$,
the adjoint operation to exterior multiplication.  This action can be used to construct
a vector space isomorphism $\sigma$
$$x\in C(V)\rightarrow \sigma(x)=x1\in\Lambda^*(V)$$
but this isomorphism doesn't respect the product operations in the two algebras.

$\Lambda^*(V)$ is not an irreducible $C(V)$ module and we now turn to the problem
of how to pick an irreducible submodule, this will be the spinor module $S$.
The way in which this is done is roughly analogous to the way in which highest
weight theory was used to pick out irreducible representations in $L^2(G)$.

\subsubsection{Complex Structures and Spinor Modules}

If one complexifies $V$ and works with $V_{\mathbf C}=V\otimes
\mathbf C$, $C(V_{\mathbf C})$ is just the complexification
$C(V)\otimes\mathbf C$ and is isomorphic to the matrix algebra of
$2^n\times 2^n$ complex matrices. $C(V_\mathbf C)$ thus has a
single irreducible module and this spinor module $S$ will be a
$2^n$-dimensional complex vector space. Knowledge of $C(V_{\mathbf
C})=End(S)$ only canonically determines $P(S)$, the
projectivization of $S$. To explicitly construct the spinor module
$S$ requires some extra structure, one way to do this begins with
the choice of an orthogonal complex structure $J$ on the
underlying real vector space $V$.  This will be a linear map
$J:V\rightarrow V$ such that $J^2=-1$ and $J$ preserves the inner
product
$$(Jv_1,Jv_2)=(v_1,v_2)$$
If $V$ is given a complex structure in this way, it then has a positive-definite Hermitian inner
product
$$(v_1,v_2)_J=(v_1,v_2) + i(v_1,Jv_2)$$
which induces the same norm on $V$ since $(v,v)_J=(v,v)$.

Any such $J$ extends to a complex linear map $$J:V_{\mathbf C}\rightarrow V_{\mathbf C}$$
on the complexification of $V$ with
eigenvalues $+i,-i$ and the corresponding eigenspace decomposition
$$V_{\mathbf C}=V^+_J\oplus V^-_J$$
$V^{\pm}_J$ will be $n$-complex dimensional vector spaces on which $J$ acts by multiplication by $\pm i$.
They are can be mapped into each other by the anti-linear conjugation map
$$v^+\in V^+_J\rightarrow \overline {v^+}\in V^-_J$$
which acts by conjugation of the complex scalars.

One can explicitly identify $V$ and $V^+_J\subset V_{\mathbf C}$ with the map
$$v \rightarrow v^+=\frac{1}{\sqrt{2}}(1-iJ)v$$
This map is an isometry if one uses the inner product $(\cdot,\cdot)_J$ on $V$ and the
restriction to $V^+_J$ of the inner
product $<\cdot,\cdot>$ on $V_{\mathbf C}$ that is the sesquilinear extension of $(\cdot,\cdot)$ on $V$
(i.e. $<v_1,v_2>=(v_1,\overline v_2)$).

The space ${\mathcal J}(V)$ of such $J$ is isomorphic to
$O(2n)/U(n)$ since each $J$ is in $O(2n)$ and a $U(n)$ subgroup
preserves the complex structure. Note that $V^+_J$ and $V^-_J$
are isotropic subspaces (for the bilinear form $(\cdot,\cdot)$)
since for $v^+\in V^+_J$
$$(v^+,v^+)=(Jv^+,Jv^+)=(iv^+,iv^+)=-(v^+,v^+)$$
The set of $J$'s can be identified with the set of maximal dimension isotropic subspaces of $V_{\mathbf C}$.

Given such a complex structure, for each $v\in V_{\mathbf C}$, one
can decompose
$$v=c(v)+a(v)$$
where
the \lq\lq creation operator" $c(v)$ is Clifford multiplication by $$c(v)=\frac{1}{\sqrt{2}}v^+=\frac{1}{2}(1-iJ)v$$
and the \lq\lq annihilation operator" $a(v)$ is Clifford multiplication by
$$a(v)=\frac{1}{\sqrt{2}}v^-=\frac{1}{2}(1+iJ)v$$
If one requires that one's representation of the Clifford algebra
is such that $v\in V$ acts by self-adjoint operators then one has
$$v^+=\overline {v^-},\ \  v^-=\overline{v^+}$$

In this case the defining relations of the complexified Clifford
algebra
$$v_1v_2+v_2v_1=2(v_1,v_2)$$
become the so-called
Canonical Anti-commutation Relations (CAR)
$$c(v_1)c(v_2)+c(v_2)c(v_1)=0$$
$$a(v_1)a(v_2)+a(v_2)a(v_1)=0$$
$$c(v_1)a(v_2)+a(v_2)c(v_1)=(v^+_1,\overline{v^+_2})=<v^+_1,v^+_2>=(v_1,v_2)_J$$

The CAR are well-known to have an irreducible representation on a
vector space $S$ constructed by assuming the existence of a \lq\lq vacuum" vector
$\Omega_J\in S$ satisfying
$$a(v)\Omega_J=0\ \  \forall v\in V$$ and applying products of
creation operators to $\Omega_J$.  This representation is
isomorphic to $\Lambda^*(V^+)$ with $\Omega_J$ corresponding to
$1\in \Lambda^*(V^+_J)$. It is $\mathbf Z_2$-graded, with
$$ S^+=\Lambda^{even}(V^+_J),\ \ S^-=\Lambda^{odd}(V^+_J)$$

\subsubsection{Spinor Representations and the Pfaffian Line Bundle}
\label{spinor-geometry}
We now have an irreducible module $S$ for $C(V_\mathbf C)$, and
most discussions of the spinor module stop at this point. We would
like to investigate how the construction depends on the choice of
$J\in {\mathcal J}(V)=O(2n)/U(n)$.  One way of thinking of this is
to consider the trivial bundle
\begin{equation*}
\begin{CD}
{} @. O(2n)/U(n)\times S\\
@. @VVV\\
{} @. O(2n)/U(n)
\end{CD}
\end{equation*}
The subbundle of vacuum vectors $\Omega_J$ is a non-trivial
complex line bundle we will call (following \cite{PS}) $Pf$.

Associating $V^+_J$ to $J$, the space ${\mathcal J}(V)$ can
equivalently be described as the space of n-dimensional complex
isotropic subspaces of $V_{\mathbf C}$ and is thus also known as
the \lq\lq isotropic Grassmannian". It is an $\frac{n(n-1)}{2}$
complex dimensional algebraic subvariety of the Grassmannian
$Gr(n,2n)$ of complex n-planes in 2n complex dimensions.
${\mathcal J}(V)$ has two connected components, each of which can
be identified with $SO(2n)/U(n)$.

${\mathcal J}(V)$ also has a holomorphic embedding in the space
$P(S)$ given by mapping $J$ to the complex line $\Omega_J\subset
S$. Elements of $S$ that correspond to $J$'s in this way are
called \lq\lq pure spinors".  To any spinor $\psi$ one can
associate the isotropic subspace of $v\in V_{\mathbf C}$ such that
$$v\cdot\psi=0$$
and pure spinors are those for which the dimension of this space
is maximal. Pure spinors lie either in $S^+$ or $S^-$, so one
component of ${\mathcal J}(V)$ lies in $P(S^+)$, the other in
$P(S^-)$.

There is a map from $S^*$, the dual of $S$, to holomorphic
sections of $Pf^*$, the dual bundle of $Pf$, given by restricting
an element of $S^*$ to the line $\Omega_J$ above the point
$J\in{\mathcal J}(V)$.  This turns out to be an isomorphism
\cite{PS} and one can turn this around and define $S$ as the dual
space to $\Gamma(Pf^*)$.  Besides being a module for the Clifford
algebra, this space is a representation of $Spin(2n)$, the spin
double cover of $SO(2n)$.  This is essentially identical with the
Borel-Weil description of the spinor representation of $Spin(2n)$.
$\Omega_J$ is a highest weight vector and we are looking at the
representation of $Spin(2n)$ on holomorphic sections of the homogeneous
line bundle $Pf^*$ over $Spin(2n)/\widetilde{U(n)}$  instead of
its pull-back to $Spin(2n)/T$. Here
$$\widetilde{U(n)}=\{(g,e^{i\theta})\in U(n)\times S^1:
det(g)=e^{i2\theta}\}$$ is the inverse image of $U(n)\subset
SO(2n)$ under the projection $Spin(2n)\rightarrow SO(2n)$.

For each $J$, we have an explicit model for $S$ given by
$\Lambda^*(V^+_J)$, with a distinguished vector $\Omega_J=1$. We
would like to explicitly see the action of $Spin(2n)$ on this
model for $S$.  In physical language, this is the \lq\lq Fock
space" and we are looking at \lq\lq Bogoliubov transformations".
The action of the subgroup $\widetilde{U(n)}$ of $Spin(2n)$ is the
easiest to understand since it leaves $\Omega_J$ invariant (up to
a phase).  As a $\widetilde{U(n)}$ representation
$$S=\Lambda^*(V^+_J)\times (\Lambda^{n}(V^+_J))^{\frac{1}{2}}$$
meaning that $S$ transforms as the product of the standard
exterior power representations of $U(n)$ times a scalar factor
that transforms as
$$ (g,e^{i\theta})z=e^{i\theta}z$$
i.e., the vacuum vector $\Omega_J$ transforms in this way. To see
this and to see how elements of $Spin(2n)$ not in
$\widetilde{U(n)}$ act we need to explicitly represent $Spin(2n)$
in terms of the Clifford algebra.

The group $Spin(2n)$ can be realized in terms of even, invertible
elements $g$ of the Clifford algebra.  Orthogonal transformations
$T_g\in SO(V)$ are given by the adjoint action on Clifford algebra
generators
$$gvg^{-1}=T_g(v)$$
and the adjoint action on the rest of the Clifford algebra gives
the rest of the exterior power representations of $SO(V)$.
If instead one considers the left action of $g$ on the Clifford
algebra one gets a reducible representation.  To get an irreducible
representation one needs to construct a minimal left ideal and
this is what we have done above when we used a chosen complex
structure $J$ to produce an irreducible Clifford module $S$.

The Lie algebra generator $L_{ij}$ that generates orthogonal rotations
in the $i-j$ plane corresponds to the Clifford algebra element
$\frac{1}{2}e_ie_j$ (where $e_i,\ i=1,\cdots,2n$ are basis for $V$) and satisfies
$$[\frac{1}{2}e_ie_j,v] = L_{ij}(v)$$
While $Spin(2n)$ and $SO(2n)$ have isomorphic Lie algebras, $Spin(2n)$ is
a double cover of $SO(2n)$ since
$$e^{2\pi L_{ij}}=1$$
but
$$e^{2\pi\frac{1}{2}e_ie_j} = -1$$
A maximal torus $T$ of $Spin(2n)$ is given by the $n$ copies of $U(1)$

$$ e^{\theta e_{2k-1}e_{2k}} \ \text{for} \ k=1,\cdots,n\ \ \text{and} \ \theta\in [0,2\pi]$$
corresponding to independent rotations in the $n$ 2-planes with coordinates $2k-1,2k$.
A standard choice of complex structure $J$ is given by a simultaneous $\frac{\pi}{2}$ rotation
in these planes. Using this one can show that the vector $\Omega_J$ transforms under $T$
with a weight of $\frac{1}{2}$ for each copy of $U(1)$, thus transforming as
$$(\Lambda^{n}(V^+_J))^{\frac{1}{2}}$$
under $\widetilde{U(n)}$.

\subsubsection{The Spinor Vacuum Vector as a Gaussian}
\label{spinor-vacuum-vector}

There is (\cite{PS}, Chapter 12.2) an explicit formula for how $\Omega_J$ 
varies with $J$ as an element of the complex exterior algebra description of 
the spin module $S$. 
This requires choosing a fixed $J_0$ and a corresponding 
decomposition 
$$V_{\mathbf C}=V^+_{J_0}\oplus V^-_{J_0}$$
This choice fixes a chart on a set of $J$'s containing $J_0$ with coordinate
on the chart given by the set of skew-linear maps
$$\omega: V^+_{J_0}\rightarrow V^-_{J_0}$$
(skew-linearity of the map implies that its graph is an isotropic subspace,
and thus corresponds to a $J$). The space of such maps can be identified with
$\Lambda^2(V^+_{J_0})$, the space of antisymmetric two-forms on $V^+_{J_0}$. Under
this identification $\Omega_J$ will be proportional to the vector
$$e^{\frac{1}{2}\omega}\in \Lambda^*(V^+_{J_0})$$

The construction we have given here of the spinor representation has a precise
analog in the case of the metaplectic representation. In that case there is
a similar explicit formula for the highest weight vector as a Gaussian, see
\cite{Segal2}.

\subsubsection{Structure of Clifford Algebra Modules}
\label{clifford-algebra-modules}

Let $C_k$ be the complexified Clifford algebra $C(\mathbf R^k)\otimes \mathbf C$ of
$\mathbf R^k$ and $M_k$ be the Grothendieck group of complex $\mathbf Z_2$ graded
irreducible $C_k$ modules. Then, using the inclusion
$$i:\mathbf R^k\rightarrow \mathbf R^{k+1}$$
$C_{k+1}$ modules pull-back to $C_k$ modules and one can consider
$$A_k=M_k/i^*M_{k+1}$$
the set of classes of $C_k$ modules, modulo those that come from $C_{k+1}$ modules.
$A_*$ is a graded ring, with product induced from the tensor product of Clifford
modules. It turns out \cite{ABS} that
\begin{equation*}
A_k=
\begin{cases}   
{\mathbf Z}&k\ \text{even}\\
0&k\ \text{odd}
\end{cases}
\end{equation*}
and the generator of $A_{2n}$ is the $n$-th power of the generator of $A_2$.

There is a natural graded version of the trace on $C_{2n}$, the supertrace,
which satisfies
\begin{equation*}
Str(\alpha)=
\begin{cases}
Tr_{S^+}(\alpha)-Tr_{S^-}(\alpha)& \alpha \in C_{2n}^{even}\\
0&\alpha \in C_{2n}^{odd}
\end{cases}
\end{equation*}

This supertrace is (up to a constant depending on conventions) identical with the Berezin
integral of the corresponding element of the exterior algebra under the map $\sigma$

$$Str(\alpha)\propto \int \sigma(\alpha)$$

\subsection{Kostant's Dirac Operator and a Generalization of Borel-Weil-Bott}

The Borel-Weil-Bott theorem gave a construction of irreducible
representations in terms of the Lie algebra cohomology of an explicit complex
built out of the exterior algebra $\Lambda^*(\mathfrak n_-)$.  This
involved an explicit choice of complex structure to define the
decomposition
$$ {\mathfrak g}/{\mathfrak t}\otimes{\mathbf C}={\mathfrak n}_+
\oplus {\mathfrak n}_-$$

If we can replace the use of $\Lambda^*(\mathfrak n_-)$ by the use of
spinors $S_{{\mathfrak g}/{\mathfrak t}}$ associated to the vector
space ${\mathfrak g}/{\mathfrak t}$, we will have a construction
that is independent of the choice of complex structure on ${\mathfrak g}/{\mathfrak t}$.
In addition, the fact that $\Lambda^*(\mathfrak n_-)$ and  $S_{{\mathfrak g}/{\mathfrak t}}$
differ as $T$ representations by a factor of
$$(\Lambda^n(\mathfrak n_-))^\frac{1}{2}$$
(here $n$ is the complex dimension of $\mathfrak n_-$), explains the
mysterious appearance in the Borel-Weil-Bott theorem of the weight
of $T$ which is half the sum of the positive roots.  This idea goes back to \cite{Bott2}.

The use of spinors also allows us to generalize Borel-Weil-Bott from the
case of $G/T$ to $G/H$ for an arbitrary $H$ with the same rank as $G$.
If $G/H=G_{\mathbf C}/P$ ($G_{\mathbf C}$ the complexification of $G$,
$P$ a parabolic subgroup), then $G/H$ is a complex manifold (actually
projective algebraic) and the original Borel-Weil-Bott theorem \cite{Bott1}
describes the representations of $G$ that occur in the sheaf cohomology groups
of homogeneous vector bundles over $G/H$.  Using spinors, one can extend
this \cite{GKRS} to the cases where $G/H$ is not even a complex manifold:

\begin{theorem}[Gross-Kostant-Ramond-Sternberg]
In the representation ring $R(H)$
$$V_\lambda \otimes S^+ -V_\lambda \otimes S^- = \sum _{w\in W(G,H)} sgn(w)U_{w(\lambda+\rho_G)-\rho_H}$$

Here $U_{\mu}$ is the representation of $H$ with highest weight $\mu$,
$S^{\pm}$ are the half-spin representations associated to the adjoint
action of $H$ on $\mathfrak g/\mathfrak h$, $W(G,H)$ is
the subgroup of $W(G,T)$ that maps the dominant Weyl chamber for $G$ into the
dominant Weyl chamber for $H$, and $\rho_{G},\rho_{H}$ are half the sum of the
positive roots of $G$ and $H$ respectively.
\end{theorem}

This theorem was motivated by consideration of the case $G=F_4,\ H=Spin(9)$,
but the simplest non-complex case is that of $G/H$ an even-dimensional
sphere ($G=Spin(2n+1),\ H=Spin(2n)$). In \cite{Kostant2} Kostant constructs an algebraic
Dirac operator
$$\Dirac :V_\lambda \otimes S^+\rightarrow V_\lambda \otimes S^-$$
such that
$$ Ker \Dirac =\sum_{sgn(w)=+}U_{w(\lambda+\rho_G)-\rho_H}, \ \ \
Coker \Dirac =\sum_{sgn(w)=-}U_{w(\lambda+\rho_G)-\rho_H}$$

In the special case $H=T$ \cite{Kostant3}, the theorem is just the original Lie algebra
cohomology version of the Borel-Weil-Bott theorem, but now in terms of
a complex involving Kostant's Dirac operator on spinors. As in the Lie algebra cohomology
case, one gets a proof of the Weyl character formula, in the form that as an element of $R(T)$,
$$V_\lambda=\frac{\sum _{w\in W(G,T)} sgn(w)U_{w(\lambda+\delta)-\delta}}{S^+ -S^-}$$
Taking $V_\lambda$ the trivial representation shows
$$ S^+ -S^-=\sum _{w\in W(G,T)} sgn(w)U_{w(\delta)-\delta}$$
so
$$V_\lambda=\frac{\sum _{w\in W(G,T)} sgn(w)U_{w(\lambda+\delta)-\delta}}{\sum _{w\in W(G,T)} sgn(w)U_{w(\delta)-\delta}}=
\frac{\sum _{w\in W(G,T)} sgn(w)U_{w(\lambda+\delta)}}{\sum _{w\in W(G,T)} sgn(w)U_{w(\delta)}}$$

So far we have seen three constructions of the irreducible representations of
$G$, of increasing generality.  In all cases we are using the
right action of $G$ on functions on $G$, separately treating the $T$ subgroup (using
Fourier analysis), and the remaining $\mathfrak g/\mathfrak t$ part of $\mathfrak g$.
To summarize they are:
\begin{itemize}

\item (Borel-Weil):  Consider $\mathfrak n_-$ invariants (lowest weights), transforming
under $T$ with weight $-\lambda$. For those $\lambda$ not in the dominant Weyl
chamber we get nothing, but for those in the dominant Weyl chamber this picks
out a one dimensional space in $V_\lambda^*$.  Out of $L^2(G)$ we get a single irreducible
representation
$$V_\lambda=\Gamma^{hol}(L_\lambda)$$

\item (Borel-Weil-Bott-Kostant):  Consider not just $H^0(\mathfrak n_-,V^*_\lambda)$ but
all of $H^*(\mathfrak n_-,V^*_\lambda)$.  Now we get a non-zero space
for each $\lambda$, but it will be in higher cohomology for
non-dominant $\lambda$. Considering $L^2(G)\otimes
\Lambda^*(\mathfrak n_-)$, to each $\lambda$ there will be an
irreducible representation (of highest weight $w(\lambda
+\delta)-\delta$ for some Weyl group element $w$) occurring inside
this space as a representative of a cohomology group
$H^*(G/T,{\mathcal O}(L_\lambda))$.

\item (Kostant): Consider not the Lie algebra cohomology complex
$V_\lambda\otimes\Lambda^*(\mathfrak n_-)$ but the complex
$$V_\lambda\otimes S^+\stackrel{\Dirac}\longrightarrow V_\lambda\otimes S^-$$
Unlike 1. and 2., this does not in any way use the choice of a complex
structure on $\mathfrak g/\mathfrak t$.  Weights are
organized into \lq\lq multiplets" of weights of
the form $w(\lambda +\delta)-\delta$ for a single dominant weight $\lambda$ and
different choices of Weyl group element $w$.  Each \lq\lq multiplet" corresponds
to a single irreducible representation of $G$.

\end{itemize}

In \cite{Landweber1}, Landweber shows that this third construction implies
that irreducible $G$ representations occur as the index of a specific
Dirac operator

\begin{theorem}[Landweber]
In the case $H\subset G$ of equal rank there is a geometric Dirac operator
$$L^2(G\times_H((S^+)^*\otimes U_\lambda))\stackrel{\Dirac_\lambda}\longrightarrow
L^2(G\times_H((S^-)^*\otimes U_\lambda))$$
such that
$$\text{Index}_G \Dirac = sgn(w) [V_{w(\lambda +\rho_\mathfrak h)-\rho_\mathfrak g}]$$
where $w$ is a Weyl group element such that $w(\lambda +\rho_\mathfrak h)-\rho_\mathfrak g$ is
a dominant weight.
\end{theorem}

In other words, if $U_\lambda$ is an element of the \lq\lq multiplet" of representations
of $H$ corresponding to a single irreducible representation of $G$, this $G$
representation occurs as the index of the given Dirac operator.  This is a special
case of a general phenomenon first described by Bott \cite{Bott2}: if
$$L^2(G\times_H M)\stackrel{D}\longrightarrow L^2(G\times_H N)$$
is an elliptic homogeneous differential operator on $G/H$ then
$$Index_G(D)=[L^2(G\times_H M)]-[L^2(G\times_H N)]$$
is an element of the representation ring $R(G)$ that only depends on the
$H$-representations $M$ and $N$ {\it{not}} on the specific operator $D$.

Bott's calculations were motivated by the recognition that the case
of homogeneous elliptic differential operators is a special case
of the general Atiyah-Singer index theorem.  In this case the index
theorem boils down to purely representation-theoretical calculations
and so can be easily checked.  We will now turn to the general
mathematical context of equivariant K-theory in which the index
theorem has its most natural formulation.

\section{Equivariant K-theory and Representation Theory}

In the previous section we have seen an explicit construction
of the irreducible representations of compact Lie groups.
There is a more abstract way of thinking about this construction
which we would now like to consider.  This uses the notion
of equivariant K-theory, a generalization of topological K-theory.

To a compact space $M$ one can associate a topological invariant,
$Vect(M)$, the space of isomorphism classes of vector
bundles over $M$.  This is an additive semi-group under the operation
of taking the direct sum of vector bundles and can be made into
a group in the same way that the integers can be constructed out of the
natural numbers.  One way of doing this is by taking pairs of elements
thought of as formal differences $\alpha - \beta$ and the equivalence
relation
$$\alpha - \beta \sim \alpha'-\beta' \Leftrightarrow \alpha + \beta' +\gamma = \alpha'+\beta +\gamma$$
for some element $\gamma$.  $K(M)$ is the additive group constructed in this
way
from formal differences of elements of $Vect(M)$. This is often called
the \lq\lq Grothendieck Construction" and $K(M)$ is called the Grothendieck group
of $Vect(M)$.
The tensor product of vector bundles gives a product on $K(M)$, making it
into a ring.  Using the suspension of $M$ and Bott periodicity, this definition
can be extended to that of a graded ring $K^*(M)$.  We will mostly be considering
$K^0(M)$ and denoting it $K(M)$.

The Grothendieck construction on isomorphism classes of
representations of a compact Lie group $G$ using the direct sum and tensor product
of representations gives the representation
ring of $G$, $R(G)$.  Using characters, $R(G)\otimes \mathbf C$ can be studied quite explicitly
as a ring of complex-valued functions, either conjugation-invariant on $G$, or Weyl-group
invariant on $T$.

Given a space $M$ with an action of a group $G$, one can look at $G$-vector bundles. These are vector
bundles $E$ over $M$ such that an element $g$ of $G$ maps the fiber $E_x$ above $x$ to the fiber $E_{gx}$ by
a vector space isomorphism.  The Grothendieck construction on isomorphism classes of these objects gives
the ring $K_G(M)$, the equivariant K-theory of $M$.  This specializes to $R(G)$ in the case of $M=pt.$, and
to $K(M)$ in the case of trivial $G$-action on $M$.

K-theory is a contravariant functor: to a map $f:M\rightarrow N$ pull-back of
bundles give a map $f^*:K(N)\rightarrow K(M)$.  For G-vector bundles and
G-maps this also holds in equivariant K-theory.

Given a representation of $H$ one can construct a $G$-vector bundle
over $G/H$ by the associated bundle construction and thus a map
$$R(H)\longrightarrow K_G(G/H)$$
This map has an inverse
given by associating to a $G$-vector bundle over $G/H$ its fiber over the
identity coset which is an $H$ representation.  These maps give the induction
isomorphism in equivariant K-theory
$$K_G(G/H) = R(H)$$

In our discussion of the Borel-Weil-Bott theorem we were looking at elements
of $K_G(G/T)$ corresponding to a weight or representation of $T$.
An abstract way of thinking of the Borel-Weil-Bott theorem is that it
gives a \lq\lq wrong-way" map
$$\pi_*:K_G(G/T)=R(T)\rightarrow K_G(pt.)=R(G)$$
corresponding to the map $\pi$ that takes $G/T$ to a point.
The existence of such a \lq\lq wrong-way" or \lq\lq push-forward" map is an
indication of the existence of a covariant functor \lq\lq K-homology", related
to K-theory in much the way that homology is related to cohomology.
We will see that, from the equivariant K-theory point of view, finding the irreducible
representations of a compact Lie group comes down to the problem of
understanding Poincar\'e duality in the $G$-equivariant K-theory of $G/T$.

\subsection {K-Homology}

Unfortunately the definition of K-homology seems to be much more subtle
than that of K-theory and all known definitions are difficult to work with,
especially in the equivariant context.

Much work on K-theory begins from the point of view of algebra, considering
isomorphism classes not of vector bundles, but of finitely generated
projective modules over a ring $R$.  For any ring $R$, the algebraic K-theory $K(R)$
will be the Grothendieck group of such modules.
For the case $R=C(M)$, the
continuous
functions on a compact space $M$, Swan's theorem gives an equivalence between
the algebraic K-theory $K(C(M))$ and the topologically defined $K(M)$.

In algebraic geometry the algebra and geometry are tightly linked.
Using the coordinate ring of a variety one has an algebraic K-theory of
algebraic vector bundles.  Here there is a covariant functor, a \lq\lq K-homology"
in this context given by taking the Grothendieck group of isomorphism
classes of coherent algebraic sheaves.

In the very general context of operator algebras one can define the
Kasparov K-theory of such an algebra.  This is a bivariant functor
and so includes a K-homology theory.  It can be defined even for non-commutative
algebras and is a fundamental idea in non-commutative geometry.  For the
algebra of continuous functions on a compact space $M$, Kasparov K-theory
is identical with topological
K-theory and comes with a corresponding covariant K-homology theory.
Unfortunately the objects representing classes in this K-homology seem
to be difficult to work with and not closely linked to the geometry.
Connes \cite{Connes} has remarked extensively on the importance of this K-homology
and the corresponding Poincar\'e duality for many applications in the
non-commutative geometry program.

A more concrete approach to the construction of K-homology, close
in spirit to that in the algebraic category, is to define the K-homology
of M as $K(\mathbf R^n, \mathbf R^n - M)$,
the Grothendieck group of complexes of vector bundles on $\mathbf R^n$,
exact off $M$ for some closed embedding of $M$ in $\mathbf R^n$.
One drawback of this construction is its non-intrinsic nature since
it relies on an explicit embedding.  In the equivariant case one
needs an equivariant embedding in some $G$ representation.

\subsection {Orientations and Poincar\'e Duality: Homology and Cohomology}

Putting aside for the moment the question the definition of K-homology, we
would like to consider abstractly what properties it should have.  We would
like K-theory ($K^*$) and K-homology ($K_*$) to have some of the same properties as
ordinary cohomology ($H^*$) and homology ($H_*$):

\begin {itemize}
\item $H^*(M)$ is a contravariant functor for
continuous maps, i.e. given $$f:M\rightarrow N$$
we have a map
$$f^*:H^*(N) \rightarrow H^*(M)$$
\item $H_*(M)$ is a covariant functor for proper maps (maps such that the inverse
image of a compact set is compact).

\item There is a cup product on $H^*(M)$
$$(\alpha, \beta)\in H^j(M)\otimes H^k(M)\rightarrow \alpha\cup\beta\in H^{j+k}(M)$$

\item There is a cap product
$$(\alpha, \beta)\in H^j(M)\otimes H_k(M)\rightarrow \alpha\cap\beta\in H_{k-j}(M)$$
and it makes $H_*(M)$ an $H^*(M)$ module.

\item For $N$ a subspace of $M$, there are
relative cohomology groups $H^*(M,N)$. 

\item (Alexander duality): For $M^\prime$ a smooth manifold of dimension $m^\prime$ and
$M$ a closed subspace of dimension $m$ one has
$$H_i(M)=H^{m^\prime -i}(M^\prime, M^\prime -M)$$
In particular,
we can choose a closed embedding of $M$ in $M^\prime=S^{n+m}$ for $n$ large enough and
then
$$H_i(M)=H^{n+m-i}(S^{n+m},S^{n+m}-M)$$

\item There is a distinguished class $\Omega\in H^*(S^n)=H^*(\mathbf R^n, \mathbf R^n-0)$, the generator of 
$H^n(S^n)$.

\item For $M$ a smooth manifold of dimension m, there is a distinguished class in $H_m(M)$, the fundamental class $[M]$.

\item (Poincar\'e duality): For $M$ a smooth manifold of dimension m the map
$$P.D.:\alpha \in H^*(M)\rightarrow \alpha\cap [M]\in H_{m-*}(M)$$
is an isomorphism.  The pairing $<\alpha, \beta>$ given by 
$$(\alpha, \beta)\in H^j(M)\otimes H_{j}(M)\rightarrow <\alpha, \beta>=\pi_*(\alpha\cap\beta)\in H_0(pt.)$$
where $\pi$ is the map
$$\pi:M\rightarrow pt.$$
is non-degenerate. Pairing with the fundamental class will be called an integration map
and denoted
$$\int_M \alpha = <\alpha,[M]>$$

\item When one has Poincar\'e duality, given a proper map
$$f:M\rightarrow M^\prime$$
one can construct a \lq\lq push-forward" or \lq\lq umkehrung" map
$$f^*:H^*(M)\longrightarrow H^*(M^\prime)$$
as follows
$$H^*(M)\stackrel{P.D.}\longrightarrow
H_{m-*}(M)\stackrel{f_*}\longrightarrow H_{m-*}(M^\prime)\stackrel{P.D.}
=H^{*-m+m^\prime}(M^\prime)$$
For the case $m > m^\prime$ this is just a generalization of the integration map.

\item For the case of a closed embedding 
$$i:M\rightarrow M^\prime$$
the push-forward map $i_*$ can be constructed as follows.  First consider the case
of a vector bundle 
$$\pi:E\rightarrow M$$
over $M$, with zero-section
$$i:M\rightarrow E$$
$E$ is said to be oriented if there is a class 
$$i_*(1)\in H^n(E, E-\{{\text zero-section}\})= H^n(E,E-M)$$
whose restriction to each fiber is
$\Omega\in H^n(\mathbf R^n,\mathbf R^n-0)=H^n(S^n)$.  Such a class
is called an orientation class or Thom class and if it exists, $E$ is
said to be orientable.  A manifold is said to be orientable if its
tangent bundle is. Now, given a closed embedding                                    
$$i:M\longrightarrow S^{m+n}$$
in a sphere of sufficiently large dimension, $M$ will have a
tubular neighborhood which can be identified with the normal
bundle $N$. One can use the excision
property of relative cohomology and Alexander duality to identify
$H^*(N,N-M)$ and $H^*(S^{m+n},S^{m+n}-M)=H_{m-*}$
and under this identification the fundamental class is just
$$[M]=i_*(1)\in  H^{n}(N,N-M)=H^n(S^{m+n},S^{m+n}-M)$$

The push-forward $i_*(\alpha)$of an arbitrary $\alpha\in H^*(M)$ can be constructed by using
the \lq\lq Thom isomorphism"
$$\alpha\rightarrow \pi^*(\alpha)\cup i_*(1) \in H^*(N,N-M)$$
\end{itemize}

The Thom
class $i_*(1)$ provides us with a sort of \lq\lq $\delta$-function" localized on $M$.
It allows one to relate integration over $M$ to integration over $N$ for $M$ embedded in $N$.
$$\int_M i^*(\alpha)=\int_N i_*(1)\cup \alpha$$.  

Another related application of the Thom class allows one to relate integration over $M$ to
integration over $s^{-1}(0)$, the inverse image of zero for some section $s$ of some vector 
bundle $E$ (transverse to the zero section) as follows
$$\int_{s^{-1}(0)}j^*\alpha=\int_M s^*(i_*(1))\cup \alpha$$
(Here $$j:s^{-1}(0)\rightarrow M$$
is just the inclusion map).

\subsection{Orientations and Poincar\'e Duality: K-theory and K-homology}

The properties of cohomology and homology discussed in the last sections
are also shared by K-theory and K-homology.  We have seen already that
$K(M)$ is defined as the Grothendieck group of vector bundles over $M$. Relative
K-theory groups $K(M,N)$ can be defined in terms of complexes of vector bundles,
exact on $N$, or by pairs of vector bundles, with a bundle map between them that
is an isomorphism on the fibers over $N$. $K(M)$ is a contravariant functor: 
pull-back of vector bundles induces a map $f^!$ on K-theory.
The cup product is induced from the tensor product of
vector bundles.

$K(M)$ is the degree zero part of a more general $\mathbf Z$-graded theory
$K^*(M)$ and we will often denote it by $K^0(M)$.  
A standard cohomology theory has the property

$$H^i(M)=H^{i+1}(\Sigma M), \ \ \ i\ge 1$$
where $\Sigma M$ is the suspension of $M$. 
K-theory in other degrees
is defined by making this suspension property into a definition

$$K^{-i}(M)=\widetilde K(\Sigma ^i (M))$$
(here on the right we need to use reduced K-theory, K-theory
of bundles of virtual dimension zero at a base-point).

This defines K-theory in non-positive degrees, the periodicity properties
of K-theory can be used to extend this to an integer grading.  We will 
here only consider complex K-theory, the K-theory of complex vector bundles.
There are similar K-theories built using real (KO) and quaternionic (KSp)
vector bundles.

The fundamental theorem of the subject is the Bott periodicity theorem
which says that K-theory is periodic with period 2:
$$K^{-(i+2)}(M)=K^{-i}(M)$$
and that
\begin{equation*}
K^{-i}(pt.)=
\begin{cases}
\mathbf Z &i\ \text{even}\\
0 &i\ \text{odd}
\end{cases}
\end{equation*}
unlike the case of ordinary cohomology, where a point has non-trivial cohomology only
in degree zero.

K-theory and the theory of Clifford algebras are linked together in a
very fundamental way \cite{ABS}.  For an expository account of this, see \cite{L-M}.
Recalling the discussion of section \ref{clifford-algebra-modules}, the relation between
K-theory and Clifford algebras is given by 
\begin{theorem}[Atiyah-Bott-Shapiro]
There is an isomorphism of graded rings
$$\alpha:A_*\rightarrow \sum_{i\ge 0}K^{-i}(pt.)$$
\end{theorem}

The isomorphism $\alpha$ can be explicitly constructed as follows. First note that  
$$M_{2n}=\mathbf Z \oplus \mathbf Z$$ 
with generators $[S]=[S^+\oplus S^-]$, the $\mathbf Z_2$ graded spin module and
$\widetilde S$, the spin module with opposite grading. Similarly
$$i^*M_{2n+1}=\mathbf Z$$
and is generated by $[S]+[\widetilde S]$.  Thus $$A_{2n}=\mathbf Z$$
with generator $[S]-[\widetilde S]$.  The isomorphism $\alpha$ associates
to this generator an element
$$[{\mathbf S}, {\mathbf {\widetilde S}}\; \mu]\in K(\mathbf R^{2n},\mathbf R^{2n}-0)=\widetilde K(S^{2n})=K^{-2n}(pt.)$$
where $\mathbf S$ is the product bundle $S\times \mathbf R^{2n}$ over $\mathbf R^{2n}$,
${\mathbf {\widetilde S}}$ is the product bundle $\widetilde S\times \mathbf R^{2n}$ and
$\mu$ is the tautological bundle map given at a point $v\in \mathbf R^{2n}$ by Clifford
multiplication by $v$.

This K-theory element $\Omega=[{\mathbf S}, {\mathbf {\widetilde S}}\; \mu]$ plays the
role for K-theory that the generator of $H^k(S^k)$ played in ordinary cohomology. In the
case $n=1$ it is sometime known as the \lq\lq Bott class" since multiplication by it
implements Bott periodicity.  Note that the K-theory orientation class 
constructed above has a great deal more structure than that of cohomology.  $Spin(2n)$
acts by automorphisms on $C_{2n}$ and on the vector bundle construction of
$[{\mathbf S}, {\mathbf {\widetilde S}}\; \mu]$.  

There is another purely representation theoretical
version of the Atiyah-Bott-Shapiro construction.  The representation ring
$RSpin(2n+1)$ is a subring of $RSpin(2n)$, the inclusion given by restriction
of representations.  It turns out that $RSpin(2n)$ is a free $RSpin(2n+1)$ module
with two generators which can be taken to be $1$ and $S^+$. Now since 
$$ Spin(2n+1)/Spin(2n)=S^{2n}$$
the associated bundle construction gives a map from $RSpin(2n)$ to vector bundles on $S^{2n}$
and this map defines an isomorphism
$$RSpin(2n)/RSpin(2n+1)\rightarrow K(S^{2n})$$

Restricting attention to even-dimensional manifolds $M$
(so we can just use complex K-theory), we can now do the same constructions as in cohomology, defining
for a vector bundle $E$ an orientation or Thom class $i_!(1)\in K(E,E-M)$ to be one
that restricts to $\Omega$ on the fibers.  Choosing a K-theory orientation of the tangent bundle $TM$ of a manifold
is equivalent to choosing a $Spin^c$ structure on the manifold. Note that there may be many possible $Spin^c$
structures on a manifold and thus many different K-theory orientations of the manifold.  Choosing a 
closed embedding
$$i:M\longrightarrow S^{2n+2k}$$
the normal bundle $N$ will be orientable if $TM$ and as in the cohomology case
the Thom class $i_!(1)$ provides an element
$$K(N, N-M)=K(S^{2n+2k},S^{2n+2k}-M)$$
Using relative K-theory as a definition of K-homology, this provides a K-homology fundamental class
which we will denote $\{M\}\in K_0(M)$.

This definition of the fundamental class in K-homology as a relative K-theory class is useful
for topological computations, but there is a very different looking
definition of K-homology called analytic K-homology that uses elliptic operators as cycles. 
The initial suggestion for the existence of this theory can be found in \cite{Atiyah1}, a recent
exposition is \cite{Higson-Roe} and many other applications are discussed in \cite{Connes}.
The identity
of these two different versions of K-homology is the content of the Atiyah-Singer
index theorem.  In analytic K-homology, the fundamental class of $M$
is the class of the Dirac operator.  The cap product map:
$$\cap :K^0(M)\otimes K_0(M) \rightarrow K_0(M)$$
is constructed by twisting the operator by the vector bundle $E$.  In particular,
cap product by the fundamental class $\{M\}=[\Dirac]$ gives a Poincar\'e duality map
$$[E]\in K^0(M)\stackrel{P.D.}\longrightarrow [E]\cap [\Dirac] =[\Dirac _E]\in K_0(M)$$

In analytic K-theory the pushforward map $\pi_*$ associated to the map
$$\pi: M \rightarrow pt.$$
takes the class of an elliptic operator $D$ in $K_*(M)$ to its index
$$\pi_*([D])=index\ D=[ker\ D] - [coker\ D]\in K_0(pt.)= \mathbf Z$$
The analog of the cohomology integration map $\int_M$ in K-theory is
the map
$$ [E]\in K(M)\rightarrow \pi_*([E]\cap [\Dirac]) =index\ \Dirac _E \in K_0(pt.)=\mathbf Z$$
so the K-theory integral over $M$ of a vector bundle $E$ is just the index of the Dirac
operator twisted by $E$.

For $D$ an elliptic pseudodifferential operator, the two versions of 
K-homology we have considered are related by the notion of the symbol
$\sigma (D)$ of the operator.  The symbol gives a bundle map between
bundles pulled-back to the cotangent bundle $T^*M$, and the map is
an isomorphism away from the zero section for the elliptic case. Thus
$\sigma (D)$ defines a class in $K(T^*M, T^*M-M)$ and so a class in our
topologically defined $K_0(M)$.  The symbol of the Dirac operator is the 
orientation class $i_!(1)$.  

The Atiyah-Singer index theorem tells us that one can compute the index of
an elliptic operator $D$ by computing the push-forward of 
$\sigma (D)\in K(T^*M, T^*M-M)$ to $K_0(pt.)=\mathbf Z$
They define this by using the fact that $TN$ is a bundle over $TM$ that
can be identified with $\pi^*(N)\otimes \mathbf C$ and embedding $TN$
in a large sphere $S^{2k}$. Then 
$i_!(\sigma (D))\in K(TN,TM)$ will give an element of $K(S^{2k})$ and thus
an integer by Bott periodicity. This integer will be the index of $D$.

One would like to be able to compute the index in terms of the more
familiar topological invariants of homology and cohomology, we will see later
how this can be done using the Chern character and other characteristic
classes.

\subsection {Orientations and Poincar\'e Duality: Equivariant K-theory}

Much of the previous section continues to hold when one generalizes from K-theory to equivariant K-theory.  
$K^0_G(M)$ is the Grothendieck group of equivariant vector bundles, it again is a ring using tensor product of
vector bundles.  One major difference is that now $K_G^0(pt.)$ is non-trivial, it is the representation ring
$R(G)$.  Now $K^0_G(M)$ has additional interesting algebraic structure: it is a module over $K_G^0(pt.)=R(G)$.
In order to define a corresponding equivariant homology theory via Alexander duality, one needs to use
an equivariant embedding in $V$ for $V$ some representation of $G$ of sufficiently large dimension.

Equivariant K-theory is naturally graded not by $\mathbf Z$, but by the representation ring. One gets elements
in other degrees than zero by considering $K^0(M\times V)$. The 
Atiyah-Bott-Shapiro construction can also be generalized. Instead of considering modules over
the Clifford algebra of $\mathbf R^n$ we need to consider the Clifford algebra of $V$, the ABS construction
then gives a complex of equivariant vector bundles over $V$, exact away from zero.   

In the equivariant context the Dirac operator continues to provide an equivariant fundamental class and
there is an equivariant integration map
$$\pi_!:K_G(M)\rightarrow K_G(pt.)=R(G)$$
given by the index map
$$ [E]\in K_G(M)\rightarrow index\ \Dirac _E \in K_G(pt.)=R(G)$$
since the index of an equivariant operator is a difference of representations, thus an element of $R(G)$.

If we specialize to the case $M=G/T$ we see that our integration map is just the map
$$\pi_!:K_G(G/T)=R(T)\rightarrow K_G(pt.)=R(G)$$
that takes a weight belonging to some multiplet to the corresponding representation of $G$.
$K_G(G/T)=R(T)$ is a module over $R(G)$, this module structure is the one that comes from taking
the restriction of a $G$ representation to $T$ and using multiplication in $R(T)$.  $R(T)$ is a free module 
over $R(G)$ of degree $|W(G,T)|$ and the analog of Poincar\'e duality in this case is seen in the
existence of a map
$$K_G(G/T)\rightarrow Hom_{K_G(pt.)}(K_G(G/T),K_G(pt.))$$
or
$$R(T)\rightarrow Hom_{R(G)}(R(T),R(G))$$
given by
$$\lambda\rightarrow (\mu\rightarrow index\ \Dirac_{L_\lambda\otimes L_\mu})$$

A generalization of equivariant K-theory was developed by Atiyah \cite{Atiyah2} and
more recently by Berline and Vergne \cite{BV1,BV2}.  
This generalization involves \lq\lq transverse elliptic" operators, i.e. ones whose symbol
may not be invertible in directions along $G$-orbits, but is on directions transverse to
the $G$-orbits.  It allows one to consider a sort of equivariant K-theory integration map
which along $G$-orbits is purely a matter of representation theory.
The integration map takes values not in the character ring $R(G)\otimes \mathbf C$
of conjugation invariant functions on $G$, but in its dual, distributions on $G$ that
are conjugation invariant.

\section{Classifying Spaces, Equivariant Cohomology and Representation Theory}

\subsection{Classifying Spaces and Equivariant Cohomology}

We first encountered homological methods in the context of the Lie
algebra cohomology $H^*(\mathfrak n, V)$.  There we studied the 
$\mathfrak n$-invariant part of a $U(\mathfrak n)$ module $V$ by
replacing $V$ with a resolution of $V$ by free $U(\mathfrak n)$   
modules.  This resolution is a chain complex whose cohomology is  
just $V$ in degree zero.  Taking the $\mathfrak n$-invariant part
of the resolution leads to a new chain complex whose degree zero 
cohomology is the invariant part of $V$. Now the complex may have
cohomology in other degrees leading to higher cohomology
phenomena.

In this section we will consider another analogous homological construction,
this time using topological spaces.             
Roughly this can abstractly be described as follows \cite{Moore}.
Instead of just a
vector space $V$ with a $G$ action, consider a topological space
$M$ with a $G$ action. The analog of $V$ here is the chain complex
of $M$, it is a differential module over the differential algebra  
$C_*(G)$ of chains on $G$. To get an analog of a free resolution
of $V$, consider a topological space $EG$ which is homotopically
trivial (contractible) and has a free $G$ action.  The complex of 
chains on $EG\times M$ will be our analog of a free resolution.
Equivariant cohomology will be the derived functor of taking
invariants of the $C_*(G)$ action.  In a de Rham model of cohomology we
are taking forms not just invariant under the $G$ action, but \lq\lq basic",
having no components in the direction of the  $G$ action.

With this motivation, the definition of equivariant cohomology is

\begin{define}
Given a topological space $M$ with a $G$ action, the $G$-equivariant
cohomology of $M$ is
$$H^*_G(M)=H^*(EG\times_G M)$$
The classifying space $BG$ of $G$ is the quotient space $EG/G$.
\end{define}

Here we are using cohomology with real coefficients.
Note that $H^*_G(M)$ is not the same as $(H^*(M))^G$, the $G$ invariant part of $H^*(M)$.
For $G$ compact taking $G$ invariants is an exact functor so $(H^*(M))^G=H^*(M)$.  $H^*_G(M)$ has the
property that, for a free $G$ action
$$H^*_G(M)=H^*(G/M)$$
At the other extreme, the equivariant cohomology of a point is now non-trivial: for
a compact Lie group 
$$H^*_G(pt.)=H^*(EG/G)=H^*(BG)=S(\mathfrak g^*)^G$$
the space of conjugation invariant polynomial functions on the Lie algebra
$\mathfrak g$.  Recalling our discussion of the maximal torus $T$ of $G$, these
are $W(G,T)$ invariant functions on the Lie algebra of $T$ so
$$H^*_G(pt.)={\mathbf R}[u_1,u_2,\cdots,u_l]^{W(G,T)}$$
where $u_i$ are coordinates on $\mathfrak t$ and $l$ is the rank of $G$.  The
equivariant cohomology $H^*_G(M)$ is not only a ring, but is a module over the
ring $H^*_G(pt.)$.

A simple example to keep in mind is $G=U(1)$ for which in some sense
$$EU(1)=\lim_{n\rightarrow\infty}S^{2n+1}=S^\infty$$
and 
$$BU(1)=\lim_{n\rightarrow\infty}CP^{n}=CP^\infty$$
the infinite dimensional complex projective space. The cohomology ring of
$CP^n$ is 
$$H^*(CP^n)={\mathbf R}[u]/u^{n+1}$$
so this is consistent with
$$H^*_{U(1)}(pt.)={\mathbf R}[u]$$
Note that the
classifying space $BG$ will be infinite dimensional for all compact $G$, but
often can be analyzed by taking limits of finite dimensional constructions.
We may want to consider cohomology classes with terms in indefinitely high
degrees, i.e. formal power series as well as polynomials, writing these as
$$H^{**}_{U(1)}(pt.)={\mathbf R}[[u]]$$

Just as in equivariant K-theory we have the identity
$$H^*_G(G/T)=H^*_T(pt.)$$
since
$$EG\times_G G/T=EG/T=BT$$
and more generally for $H$ a subgroup of $G$ there is an induction relation
$$H^*_H(M)=H^*_G(G\times_H M)$$

If one wants to work with an explicit de Rham model of equivariant cohomology,
one would like to avoid working with differential forms on the infinite 
dimensional space $EG$.  In an analogous
fashion to what one does when one computes de Rham cohomology of
compact groups, where by an averaging argument one replaces forms
on $G$ by left-invariant forms which are generated by
left-invariant 1-forms (the Lie algebra $\mathfrak g$), one can use a much smaller
complex, replacing $\Omega^*(EG)$ by the \lq\lq equivariant differential
forms"
$$\Omega_G^*(M)=\{W(\mathfrak g)\otimes \Omega^*(M)\}_{basic}=
(S({\mathfrak g}^*)\otimes \Omega^*(M))^G$$

Here 
$$W({\mathfrak {g}})=S({\mathfrak g}^*)\otimes\Lambda({\mathfrak g}^*)$$ 
is the Weil algebra of $G$, a finite dimensional model for forms on $EG$.
For more details on equivariant cohomology see \cite{Atiyah-Bott} and \cite{Ginzburg}
and for
a detailed exposition of the formalism of equivariant forms and the
equivariant differential, see \cite{GS}.  There one can also find
a translation of this formalism into the language of superalgebras.  In
this language $\Omega_G^*(M)$ is a superalgebra with action of the
Lie superalgebra
$$\tilde{\mathfrak g} ={\mathfrak g}_{-1} \oplus {\mathfrak g}_0 \oplus {\mathfrak g}_1$$
where ${\mathfrak g}_0=\mathfrak g$, the ${\mathfrak g}_{-1}$ is a copy of $\mathfrak g$
corresponding to the interior products in the $\mathfrak g$ directions and ${\mathfrak g}_1$
is one dimensional, corresponding to the differential $d$.  Taking basic forms
is then equivalent to $\tilde{\mathfrak g}$ invariance, roughly corresponding to the
motivation of taking $C_*(G)$ invariants given at the beginning of this section.

\subsection{Classifying Spaces and Equivariant K-Theory}

For a general group $G$ the topology of the classifying space of a
group $G$ is often closely related to the representation theory of
$G$. Given a representation $V$ of $G$, one can form the vector
bundle $$V_G=EG \times _G V$$over $BG$. In terms of equivariant
K-theory classes one has a map 
$$V\in K_G(pt.)=R(G)\rightarrow [V_G]\in K_G(EG)=K(BG)$$ 
For compact Lie groups
$$K_G(EG)=K(BG)=\widehat {R}(G)$$ where $\widehat {R}(G)$ is the
completion of the representation ring at the ideal of virtual
representations of zero dimension \cite{AH}.  Thinking of $R(G)$ as the ring
of conjugation invariant functions on $G$ in this case our map
$$V\in R(G)\rightarrow \widehat{R}(G)$$ is just the map that takes a global
function to its power series expansion around the identity.

As an explicit example, consider the case $G=U(1)$. Then the representation ring is
$$R(U(1))=\mathbf Z [t, t^{-1}]$$
Writing a representation $t$ in terms of its character $\chi(t)$ as a function on 
${\mathfrak g}=\mathbf R$
$$\chi(t)=e^u$$
Then 
$(R(U(1))\otimes \mathbf C)$ the subring of the power series ring ${\mathbf C}[[u]]$
generated by $\{e^u,e^{-u}\}$ and the completion at $e^u=1$ is the power series ring
$$\widehat{R}(U(1))\otimes \mathbf C={\mathbf C}[[z]]$$
where $z=e^u-1$.

A lesson of this is that one can study the character ring
$R(G)$ by considering the
equivariant K-theory of $EG$, although at the cost of
only being able to work with power series expansions of characters on the
Lie algebra instead of global character functions. Furthermore, when one
does this one ends up with something that is equivalent to equivariant cohomology.
 
In some sense the map
$$\pi_!:K_G(EG)\rightarrow K_G(pt.)$$
while only being a homotopy equivalence, not an equivariant homotopy
equivalence, still closely relates these two rings.  
The Baum-Connes conjecture implies that something like this phenomenon occurs even in the
context of discrete groups $G$. More specifically they conjecture \cite{BCH} that
$$K_{G*}(\overline{EG})=K_*(C_r^*(G))$$
Here $\overline{EG}$ is their classifying space for proper G
actions, on the left one is using some version of K-homology,
typically one defined in terms of operator algebras, such as
Kasparov KK-theory. On the right is the operator algebra K-theory
of the reduced $C^*$-algebra of G.

\subsection {Equivariant Homology and the Fundamental Class}

An obvious possible integration map in equivariant cohomology is
integration of differential forms which gives a map
$$\int_M:H^*_G(M)\rightarrow H^*_G(pt.)=S(\mathfrak g^*)^G$$
We can use to this to construct a map similar to the one we saw in equivariant K-theory reflecting
Poincar\'e duality
$$H^*_G(M)\rightarrow Hom _{S(\mathfrak g^*)^G}(H^*_G(M),S(\mathfrak g^*)^G)$$
defined by
$$[\alpha]\rightarrow ([\beta]\rightarrow \int_M \alpha\beta)$$
This map is an isomorphism in certain cases \cite{Ginzburg}, for instance
if $M$ is a compact symplectic manifold and the $G$ action is Hamiltonian (preserves
the symplectic form).  In other cases it fails to be an isomorphism, most dramatically
in the case of a free $G$ action when the map is zero since the integral of an equivariantly
closed form vanishes.  

This failure is an indication that to get a well-behaved notion of an
equivariant homology theory dual to equivariant cohomology in terms of differential
forms, one needs to consider not just $S(\mathfrak g^*)$, the polynomial functions on $\mathfrak g$, but the
class of generalized functions on $\mathfrak g$, $C^{-\infty}(G)$.  We'll begin by considering the
case of a point.

\subsubsection{The Equivariant Homology of a Point}

In equivariant K-theory,
$K_G(pt.)\otimes {\mathbf C}=R(G)\otimes {\mathbf C}$ is the space of 
character functions on $G$.  There is a natural inner product given by
the Haar integral over $G$, and different irreducible representations $V$ and $W$ are
orthonormal with respect to this inner product. 
\begin{equation*}
<\chi_V,\chi_W>=\int_G \chi_V(g)\overline{\chi_W}(g)dg=
\begin{cases}
1&V\simeq W\\
0&V\not\simeq W
\end{cases}
\end{equation*}
The irreducible representations thus give a distinguished basis for $K_G(pt.)$.
For any representation $V$, irreducible or not, its decomposition into
irreducibles (coordinates with respect to the distinguished basis) 
can be found by computing integrals.  The integral
$$\int_G\chi_V(g)\overline{\chi_{V_i}(g)}dg$$
gives the multiplicity $mult(V_i,V)$ of the irreducible $V_i$ in $V$. In particular
$$\int_G\chi_V(g)dg$$
gives the multiplicity $mult(1, V)$ of the trivial representation in $V$.  For each
irreducible $V_i$, the multiplicity $mult(V_i,\cdot)$ is a linear functional
on $K_G(pt.)$ giving a distinguished basis for the dual space to $K_G(pt.)$.  
Another important linear functional is given by evaluating the character
at the identity, giving the virtual dimension of an element of $K_G(pt.)$.

In equivariant cohomology $H^{**}_G(pt.)$ consists of power series about
zero in the Lie algebra. The (topological) dual vector space is the space
$$C^{-\infty}(\mathfrak g)_0$$
of distributions supported at $0$.  Fourier transformation maps $S^*(\mathfrak g)$ 
(polynomials on ${\mathfrak g}^*$) to distributions on $\mathfrak g$ supported at $0$: derivatives
of the $\delta$-function.  We would like to be able to define analogs of the
linear functionals $mult(V_i,\cdot)$, for instance $mult(1,\cdot)$, the multiplicity of the
identity representation.  This is not so easily defined as in the $R(G)$ case, but
one way of doing this is to compute limits of ratios such as the following:

$$\oint_{\mathfrak g} f(u)\equiv\lim_{\epsilon\rightarrow 0} \frac{\int_{\mathfrak g} f(u) e^{-\epsilon ||u||^2}} 
{\int_{\mathfrak g}e^{-\epsilon ||u||^2}}$$
where $||\cdot||$ is a conjugation invariant norm on $\mathfrak g$.

One could ask what the fundamental class is in the equivariant homology of a point.  The fact
that $H^*_G(pt)=H^*(BG)$ would indicate that in some sense the answer should be the homology fundamental
class of $BG$.  The infinite dimensionality of $BG$ makes this very ill defined, a better answer
to this question would be the functional $\oint_{\mathfrak g}(\cdot)$.

\subsubsection{The Case of a Free Action}

We saw at the beginning of this section that for a free action the obvious
integration map $\int_M$ on $H^*_G(M)$ gives zero.  To get an integration
map with better properties, we proceed much as in the case of the cohomology
Thom class. There one could replace integration over $M$ by integration
over some $N$ in which $M$ is embedded.  
The analog of the Thom class in this
case is something we will call the Witten-Thom class \cite{Witten-localization} and to define it we need
to work with a generalization of equivariant differential forms.  For a 
definition of the complex $C^{-\infty}({\mathfrak g},\Omega^*(M))$ of equivariant differential
forms with generalized coefficients and the corresponding generalization $H^{-\infty}_G(M)$, as
well as more details of this construction, see 
\cite{Kumar-Vergne}.  For a related discussion of the fundamental homology class for a free action, 
see \cite{Austin-Braam}.

\begin{define}
Given a fibration
\begin {equation*}
\begin{CD}
G @>>>P \\
@. @VV\pi V\\
{} @. M
\end{CD}
\end{equation*}
with a connection $\omega$ on $P$, define a one-form on $P\times \mathfrak g^*$
by
$$\lambda(p,u)=\omega _p(u)$$
where $\omega _p$ is the $\mathfrak g$-valued connection 1-form at $p\in P$ and 
$\omega _p(u)$ is its evaluation on the element $u\in \mathfrak g^*$. The Witten-Thom
class $\gamma \in C^{-\infty}({\mathfrak g},\Omega^*(P))$ is defined by
$$\gamma=\frac{1}{(2\pi)^{dim\ \mathfrak g}}\int_{{\mathfrak g}^*}e^{id_{\mathfrak g}\lambda}$$
where $d_{\mathfrak g}$ is the equivariant differential in $C^{-\infty}({\mathfrak g},\Omega^*(P))$.
\end{define}

The Witten-Thom class has the following properties:
\begin{itemize}
\item
Its class in $H^{-\infty}_G(P)$ is independent of the connection $\omega$ chosen to construct it.
\item
(Analog of Thom isomorphism): The map
$$ H^*(M)\rightarrow H^{-\infty}_G(P)$$
on cohomology induced from the map
$$\alpha\rightarrow \pi^*(\alpha)\wedge\gamma$$
is an isomorphism of $C^\infty(\mathfrak g)^G$ modules.
\item Evaluation of $\gamma$ on $\phi\in (C^\infty(\mathfrak g))^G$ gives
$$\int_{\mathfrak g} \gamma (u)\phi(u)= v_\omega \wedge \phi(\Omega)$$
i.e.
$$\gamma(u)=v_\omega\wedge\delta(u-\Omega)$$
Here $\Omega$ is the curvature of $\omega$ and $v_\omega$ is a vertical form on
$P$ of degree $dim\ G$ and integral $vol(G)$ over each fiber.
\item
Integration over $P$ and over $M$ are now related by
$$\int_{\mathfrak g}\phi (u)\int_P \gamma\wedge\pi^*\alpha = vol(G)\int_M \alpha\wedge\phi(\Omega)$$
For $\alpha=1$, this formula gives an expression for the characteristic number 
$$\int_M \phi(\Omega)$$
of $P$ corresponding
to $\phi$.  Taking $\phi$ of the form
$$\phi(u)=e^{-\epsilon ||u||^2}$$
and taking the limit as $\epsilon$ goes to zero
gives the formula
$$\oint_{\mathfrak g} \int_P \gamma\wedge\pi^*\alpha = \int_M \alpha$$
\end{itemize}

The last of these properties shows that we have an integration map on $H^*_G(P)$ for the case of a free
action is given by 
$$\alpha\in H^*_G(P) \rightarrow \frac{1}{vol(G)}\oint_{\mathfrak g}\int_P \gamma\wedge\alpha \in H^*(pt)$$.

\subsection{K-theory and (Co)homology: The Chern Character}

We have seen that for a point, the relation between equivariant K-theory and equivariant
cohomology is just that between a representation $V$ and its character
$$\chi_V=Tr_V(e^u)$$
expressed as a power series about $0\in\mathfrak g$.  In general one would like to
relate K-theory calculations to more familiar cohomological ones, this is done with
a map called the Chern character.

In ordinary K-theory and cohomology, the Chern character is a map
$$ch:K(M)\rightarrow H^*(M,\mathbf C)$$
and is a ring isomorphism. It takes the direct sum of vector bundles to the sum
of cohomology classes and the tensor product of vector bundles to the cup product
of cohomology classes and takes values in even dimensional cohomology. A generalization to
the odd K-theory groups takes values in odd dimensional cohomology.

Chern-Weil theory gives a de Rham cohomology representative
of the Chern character map using the curvature of an arbitrarily chosen connection.
If an element $[E]$ of K-theory is the associated bundle to a principal bundle $P$
of the form 
$$E=P\times _G V$$
for a representation $V$ of $G$, then the Chern-Weil version of the Chern character
is defined by
$$\int_{\mathfrak g} \gamma (u)\phi(u)= v_\omega \wedge \phi(\Omega)=v_\omega \wedge ch([E])$$
where $\gamma$ is the Thom-Witten form and 
$$\phi(u)=Tr_V(e^u)\in C^\infty(\mathfrak g)$$

One can define similarly define an equivariant Chern character map
$$ch_G:K_G(M)\rightarrow H^*_G(M)$$
a Chern-Weil version of this uses $G$-invariant connections.
It is no longer a rational isomorphism since as we have seen even for a point
one has to take a completion of $K_G$.  Several authors have defined a
\lq\lq de-localized equivariant cohomology" with the goal of having a 
Chern-character that is an isomorphism. For one version of this, see \cite{Duflo-Vergne}.

Quillen \cite{Quillen} has defined a generalization of the Chern character to
the case of relative K-theory, using a generalization of Chern-Weil theory that
uses \lq\lq superconnections" instead of connections.  Whereas a connection on
a $\mathbf Z_2$-graded vector bundle preserves the grading, a superconnection has
components that mix the odd and even pieces. He constructs an explicit Chern
character map that uses the superconnection
$$d+t\mu$$ on the trivial spin bundles $\mathbf S$ and $\mathbf {\widetilde S}$ over $\mathbf R^{2n}$
and takes the K-theory orientation 
class
$$[{\mathbf S}, {\mathbf {\widetilde S}}\; \mu]\in K(\mathbf R^{2n}, \mathbf R^{2n}-0)$$
to an element 
$$ Str( e^{(d+t\mu)^2}) \in H^*(\mathbf R^{2n}, \mathbf R^{2n}-0)$$
Here $Str$ is the supertrace, $t$ is a parameter, and this differential form becomes more and more
peaked at $0$ as $t$ goes to infinity.

This construction is generalized in \cite{Mathai-Quillen} to an equivariant one for
the group $G$, where $G$ is some subgroup of $Spin(2n)$. The Atiyah-Bott-Shapiro
construction of the orientation class is $G$-equivariant and gives an element in
$K_G(\mathbf R^{2n}, \mathbf R^{2n}-0)$. Mathai-Quillen use an equivariant 
superconnection formalism to map this to an equivariant differential form that 
represents an element of $H_G(\mathbf R^{2n}, \mathbf R^{2n}-0)$.
Given such an equivariant differential form, for any vector bundle $E$ over $M$ with connection of the form
$$P\times_G V$$
for some $G$-representation $V$, one can use Chern-Weil theory to get an explicit representative
for the Chern character of the K-theory orientation class
$$ch(i_!(1))\in H^*(E,E-M)$$ 
Mathai and Quillen show that this is not identical with the cohomology orientation class, 
instead it satisfies
$$ch(i_!(1))=\pi^*(\hat A(E)^{-1})i_*(1)$$
where $\hat A(E)$ is the characteristic class of the vector bundle $E$ corresponding to the function
$$\phi(u)=det^{\frac{1}{2}} (\frac{u/2}{e^{u/2}-e^{-u/2}})$$ 

The Chern character can be used to get an explicit cohomological form of the index theorem.
The index of an operator $D$ with symbol $\sigma(D)$ will be given by evaluating 
$$ch(i_!(\sigma))$$
on the fundamental class of $TN$, where $i:TM\rightarrow TN$ is the zero-section.  Standard
manipulations of characteristic classes give the well-know formula 
$$index \Dirac _E =\int_M ch(E)\wedge \hat A(TM)$$

In the equivariant context, Berline and Vergne \cite{BV0,BGV} found a generalization of this formula  
expressing the equivariant index (as a power series in $H_G^*(pt.)$) as an integral over equivariant
versions of the Chern and $\hat A$ classes.  They note that in the special case of $M=G/T$ a co-adjoint
orbit, their formula reduces to an integral formula for the character due to Kirillov.  See 
\cite{BV1,BV2,Duflo-Vergne} for generalizations to the case of transverse elliptic operators and
to de-localized equivariant cohomology.

\section{Witten Localization and Quantization}
\label{witten-localization}

Given a geometric construction of a representation $V$, a central problem
is to understand its decomposition into irreducible representations $V_i$. 
Since
$$mult(V_i,V)=mult(1,V\otimes V_i^*)$$
once one knows the irreducible representations, one just needs to be able
to compute the multiplicity of the identity representation in an arbitrary
one.  For a compact Lie group, if one knows the character $\chi _V$, this is
just the integral over the group. If the character is only known as a power
series expansion in the neighborhood of the identity, we have seen that one
still may be able to extract the multiplicity from an integral over the Lie
algebra.

Witten in \cite{Witten-localization} considered integrals of this form
for integrands such as the equivariant Chern character that occurs in the
Berline-Vergne-Kirillov integral formula for an equivariant index.  Given
a quantizable symplectic manifold $M$ with a Hamiltonian $G$ action, one
considers an equivariant line bundle $L$ which gives a class
$$[L]\in K_G(M)$$
As an element of $R(G)$, the quantization here is the K-theory push-forward
$$\pi_!([L])\in K_G(pt.)=R(G)$$
and one would like to know how this decomposes into irreducibles.  This
requires evaluating integrals of the form
$$\oint_{\mathfrak g}\int_M ch_G([L])\hat A_G(M)$$
The equivariant Chern character is represented by an equivariant differential
form that is the exponential of the equivariant curvature form of $L$
$$ch_G([L])=e^{\omega +i2\pi<\mu,u>}$$
where $\omega$ is the symplectic form (curvature of $L$) and $\mu$ is a the
moment map
$$\mu:M\rightarrow {\mathfrak g}^*$$

The integral over $\mathfrak g$ has exponential terms of the form
$$\int_{\mathfrak g}e^{-\epsilon ||u||^2 +i2\pi<\mu,u>} (\cdots)$$
and as $\epsilon$ goes to zero the integral will be dominated by contributions from
a neighborhood of the subspace of $M$ where $\mu=0$.  The Witten-Thom form relates
such integrals to integrals on $\mu^{-1}(0)/G$.  Various authors
have used this sort of principle to show that \lq\lq quantization commutes with
reduction", i.e. the multiplicity of the trivial representation in $\pi_!([L])$
is the dimension one would get by quantizing the Marsden-Weinstein reduced
phase space $\mu^{-1}(0)/G$.  For a review of some of these results, see
\cite{Sjamaar}.

\section{Geometrical Structures and Their Automorphism Groups}

\subsection{Principal Bundles and Generalized G-structures}

A principal $G$-bundle $P$
\begin {equation*}
\begin{CD}
G @>>>P \\
@. @VV\pi V\\
{} @. M
\end{CD}
\end{equation*}
can be thought of as a generalization of a group to a family of identical groups 
parametrized by a base-space $M$.  We've already seen two classes of examples: $G$ is a principal $H$-bundle
over $G/H$, and $EG$ is a principal $G$-bundle over $BG$.  This second example is universal: every
$G$ bundle $P$ occurs as the pullback under some map $f:M\rightarrow BG$ of $EG$.

To any n-dimensional manifold $M$ one can associate a principal $GL(n,{\mathbf R})$ bundle $F(M)$, the bundle whose
fiber above $x\in M$ is the set of linear frames on the tangent space $T_xM$.  A sub-bundle of
$P\subset F(M)$ with fiber $G\subset GL(n,{\mathbf R})$ will be called a $G$-structure on $M$.  
The frame bundle $F(M)$ or a 
$G$-structure $P$ can be characterized by the existence of a tautologically defined equivariant form
$$\theta \in \Omega^1(P)\otimes \mathbf R^n$$
such that $\theta _p (v)$ gives the coordinates of the vector $\pi _* (v)$ with respect to the frame $p$.

This classical definition of a $G$-structure is not general enough to deal even with the spinor geometry
of $M$ since one needs to consider a structure group $Spin(n)$ or $Spin^c(n)$ at each point and
these are not subgroups of $GL(n,{\mathbf R})$. To get
a sufficient degree of generality, we'll define

\begin{define}
A generalized $G$ structure on a manifold $M$ is a principal $G$-bundle $P$ over $M$ and a
representation 
$$\rho:G\rightarrow GL(n,\mathbf R)$$
 together with a
$G$-equivariant horizontal ${\mathbf R}^n$-valued one-form $\theta$ on $P$.
\end{define}

Note that $\rho$ does not have to be a faithful representation, it may have a kernel which can
be thought of as an \lq\lq internal symmetry".

\subsection{The Automorphism Group of a Bundle}

To any principal bundle $P$ one can associate a group, the group $Aut P$ of automorphisms of the bundle.
This group will be infinite-dimensional except in trivial cases.

\begin{define}
$Aut (P)$=\{$f:P\rightarrow P$ such that $f(pg)=f(p)g$)\}
\end{define}

$Aut (P)$ has a subgroup consisting of vertical automorphisms of $P$, those which take a point $p\in P$ to another point
in the same fiber. This is the gauge group ${\mathcal G}_P$ and elements of ${\mathcal G}_P$ can be given as maps
$$h: P\rightarrow G$$
here $h(p)$ is the group element such that
$$f(p)=ph(p)$$

The condition that $f(p)\in {\mathcal G}_P$ implies that the functions $h$
satisfy a condition $h(pg)=g^{-1}h(p)g$ so
\begin{define}
${\mathcal G}_P$=\{$h:P\rightarrow G$ such that $h(pg)=g^{-1}h(p)g$\}
\end{define}
in other words ${\mathcal G}_P$ consists of the space of sections $\Gamma (Ad P)$ where $Ad P$ is the bundle
$$Ad P=P\times_G G$$
associated to $P$, with fiber $G$ and the action of $G$ on itself the adjoint action.

First consider the case of a $G$ bundle over a point, in other words $P$ is $G$ itself.  Describe Aut P in this case, show that it is $G_L\times Z(G)$
May also want to include an explicit discussion of the one-dimensional case.

The gauge group ${\mathcal G}_P$ is a normal subgroup of $Aut(P)$ and the
quotient group is the group $Diff(M)$ of diffeomorphisms of the base space $M$.  So we have an exact sequence 
$$1\rightarrow {\mathcal G}_P\rightarrow Aut(P)\rightarrow Diff(M)\rightarrow 1$$

Taking infinitesimal automorphisms we have the exact sequence
$$0\rightarrow Lie ({\mathcal G}_P)\rightarrow Lie(Aut(P))\stackrel{\pi_*}\rightarrow Vect(M)\rightarrow 0$$
Here we are looking at the Lie algebra of $G$-invariant vector fields
on $P$, it has a Lie subalgebra of vertical $G$-invariant vector fields and a quotient Lie algebra of $G$-invariant vector field modulo vertical
ones, which can be identified with $VF(M)$, the vector fields on $M$.  The maps in this exact sequence
are both Lie algebra homomorphisms and homomorphisms of $C^\infty(M)$ modules.

\subsection {Connections}

The fundamental geometrical structure that lives on a bundle and governs how the fibers are related is called a connection.  
There are many equivalent ways of defining a connection, perhaps the simplest is

\begin{define}
A connection is a splitting $h$ of the sequence
$$0\rightarrow Lie ({\mathcal G}_P)\rightarrow Lie(Aut(P))\stackrel{\pi_*}\rightarrow VF(M)\rightarrow 0$$
i.e. a homomorphism 
$$h:VF(M)\rightarrow  Lie(Aut(P))$$
of $C^\infty(M)$ modules such that $\pi_* h=Id$.
\end{define}

For any vector field $X\in VF(M)$, $hX \in  Lie(Aut(P))$ is the horizontal lift of $X$.  A connection can
also be characterized by the distribution on $P$ of horizontal subspaces $H_p$. $H_p$ is the subspace of
$T_pP$ spanned by the horizontal lifts of vector fields.

The map $h$ is not necessarily a Lie algebra homomorphism.  When it is the connection is said to be flat. In
general the curvature $\Omega$ of a connection is characterized by the map
$$\Omega :VF(M)\times VF(M) \rightarrow Lie ({\mathcal G}_P)$$
given by
$$\Omega (X,Y)=[hX,hY]-h[X,Y]$$

A connection can dually be characterized by an equivariant $\mathfrak g$ valued one-form 
$$\omega \in \Omega^1(P)\otimes\mathfrak g$$
which satisfies
$$H_p=ker \ \omega _p$$

The space ${\mathcal A}_P$ is an affine space, the difference of two connections is a one-form
on $M$ with values in $\Gamma (Lie (Ad P))$.  The group $Aut(P)$ acts on ${\mathcal A}_P$ by
pull-back of connection one-forms.

\section{Quantum Field Theories and Representation Theory: Speculative Analogies}

We began this article with some fundamental questions concerning the significance of the
mathematical structures appearing in the quantum field theory of the Standard Model. Here
we would like to provide some tentative answers to these questions by speculating on
how these mathematical structures fit into the representation theoretical framework
we have developed in previous sections.

\subsection{Gauge Theory and the Classifying Space of the Gauge Group}

Given an arbitrary principal bundle $P$, we have seen that its automorphism group
$Aut(P)$ has a normal subgroup ${\mathcal G}_P$ of gauge transformations which acts
on ${\mathcal A}_P$, the space of connections on $P$.  ${\mathcal G}_P$ has a subgroup
${\mathcal G}^0_P$ of based gauge transformations, those that are the identity on a fixed
fiber in $P$.  ${\mathcal A}_P$ is contractible
and the ${\mathcal G}^0_P$ action is free, so we have
$$E{\mathcal G}^0_P={\mathcal A}_P,\ \ \ B{\mathcal G}^0_P={\mathcal A}_P/{\mathcal G}^0_P$$

We have seen in the case of a compact Lie group $G$ that the use of the classifying
space $BG$ allows us to translate problems about the representation theory of $G$ into
problems about the topology of $BG$, which are then studied using K-theory or (co)homology.
To any representation $V$ of $G$ is associated an element $[V_G]\in K_G(EG)=K(BG)$ and
$K_G^0(EG)=K^0(BG)$ is $\widehat {R(G)}$, the representation ring of $G$, completed at the
identity.  In the case of the gauge group ${\mathcal G}_P$, our lack of understanding of
what $R({\mathcal G}_P)$ might be is profound, but perhaps quantum field theory is telling
us that it can be approached through the study of the topological functors
$K^*_{{\mathcal G}_P}(E{\mathcal G}_P)$ and $H^*_{{\mathcal G}_P}(E{\mathcal G}_P)$.  The infinite 
dimensionality of
${\mathcal G}_P$ makes the study of these functors difficult, but if physicist's path
integral calculations can be interpreted as formal calculations involving these rings,
then there is a large amount of lore about what can be sensibly calculated that will become
accessible.

Following this line of thought, the Standard Model quantum field theory path integral
would involve a de Rham model
$$\Omega^*_G({\mathcal A}_P)=\{W(Lie({\mathcal G}_P)\otimes \Omega^*({\mathcal A}_P)\}_{\text basic}$$
of the equivariant cohomology of ${\mathcal A}_P$, and elements of
the path integral integrand may come from this de Rham model. One
obvious objection to this program is that this is what is done in
Witten's \lq\lq topological quantum field theory" (TQFT)
formulation of Donaldson invariants \cite{Witten-TQFT} (for a
review from this point of view, see \cite{Cordes-Moore-Ramgoolam})
and our physical theory should have observables that are not
topological invariants.  But perhaps the answer is that here one
is doing \lq\lq equivariant topological quantum field theory" and so has
observables corresponding to the infinitesimal actions of all
symmetries one is considering.  

TQFT will correspond to restricting attention to the subspace of $Aut(P)$ invariants,
this gives a finite dimensional purely topological problem, corresponding physically
to choosing to only study the structure of the vacuum state of the more general
equivariant theory.

One also needs to understand why the path integral is expressed as an integral over the classifying
space $B{\mathcal G}_P={\mathcal A}_P/{\mathcal G}_P$.  In the Baum-Connes conjecture case described
in an earlier section, the conjecture amounts to the idea that the map
$$K_{*G}(EG)\stackrel{\pi_*}\rightarrow K_{*G}(pt.)$$
induced from the \lq\lq collapse" map $\pi : EG\rightarrow pt.$ is an isomorphism for $G$ a discrete group.
Assuming that something like this is still true for $G={\mathcal G}_P$, one
can perhaps interpret path integral expressions as equivariant homology classes on ${\mathcal A}_P$.  Then
the integral is a reflection of the existence of some sort of fundamental class providing via cap-product
a map
$$K^*_{{\mathcal G}_P}({\mathcal A}_P)\rightarrow K_{*{\mathcal G}_P}({\mathcal A}_P)$$
Here again the notion of an equivariant fundamental class is the critical one. Our proposal is that the
Standard Model path integral involves a ${\mathcal G}_P$-equivariant fundamental class of
${\mathcal A}_P$ and that path integrals are actually calculations in an explicit model of
equivariant cohomology related to the abstract equivariant $K$-theory picture by a Chern character map.

Finally one would like to understand the occurrence of the
exponential of the Yang-Mills action as a factor in the path
integral integrand.  This can be understood by
a generalization of the arguments discussed in Section
\ref{witten-localization}, which were
originally developed by Witten for the sake of applying them
formally to the case of connections on a principal bundle $P$ over
a Riemann surface $\Sigma$.  In that case ${\mathcal A}_P$ is a
(infinite dimensional) symplectic manifold with a line bundle $L$
(the determinant bundle for the Dirac operator) whose curvature is
the symplectic form. The whole setup is equivariant under the
gauge group ${\mathcal G}_P$ and the moment map corresponding to
the ${\mathcal G}_P$ action is given by the curvature two-form
$F_A$ of the connection $A\in{\mathcal A}_P$.  

The localization
principle shows that trying to pick out the $Lie{\mathcal G}_P$
by doing an integral 
$$\lim_{\epsilon\rightarrow 0}\int_{Lie{\mathcal G}_P}\int_{{\mathcal A}_P}e^{-\epsilon ||u||^2}ch_{{\mathcal G}_P}(L)$$
leads to integrals of the form

$$\lim_{\epsilon\rightarrow 0} \int_{{\mathcal A}_P} e^{-\frac{1}{\epsilon ^2}{||F_A||^2}}(\cdots)$$
a form which is precisely that of the standard Yang-Mills theory.  Note that the limit $\epsilon\rightarrow$ we
would like to take is the same as that of taking the coupling constant to zero in the continuum limit as
specified by the asymptotic freedom of the theory.  The fact that this limit needs to be taken in a very
specific way in order to get a non-trivial continuum limit should be a highly non-trivial idea necessary
for the construction of interesting representations of $Aut(P)$ in higher dimensions.

While this specific construction only works for a Riemann surface, for a four dimensional hyperk\"ahler
manifold there is an analogous moment map. Here the zeros of the moment map are connections satisfying
$F^+_A=0$ and a Gaussian factor in this moment map and the Yang-Mills action are related by
$$e^{-\frac{1}{g^2}||F_A||^2}=e^{-\frac{1}{g^2}(||F^+_A||^2+||F^-_A||^2)}=(const.)e^{-\frac{1}{g^2}||F^+_A||^2}$$
since 
$$||F^+_A||^2-||F^-_A||^2$$
is a constant topological invariant of the bundle $P$.

\subsection{An Analogy}

The fundamental problem faced by any attempt to pursue the ideas of the previous section is
our nearly complete ignorance of the representation theory of the group $Aut(P)$ for
base spaces of more than one dimension.  In previous sections certain techniques for
decomposing 
the space of functions on a compact Lie group $G$ into irreducibles were explained in great detail partly because many of
the same techniques may be useful in the $Aut(P)$ case.  One can begin by trying to decompose
a space of functions on $P$, say $C^\infty (P)$, as an $Aut(P)$ representation.  As such it is
highly reducible but perhaps the same homological techniques of using Clifford algebras and
classifying spaces can be of help.  

Some details of the analogy we have in mind are in the following table
 ($G_L$, $G_R$, $T_L$, $T_R$ are the right and left actions of $G$ or 
$T$ on $G$).

\begin{center}
\begin{tabular}{|c|c|}
\hline
Representations of $G$ & Representations of $Aut(P)$ \\
\hline
\hline
$G,\ C^\infty(G)$ & $P,\ C^\infty(P)$\\
\hline
$T_R$ & $G$\\
\hline
$G/T$ & $P/G=M$\\
\hline
$G_L$ & $Aut P$\\
\hline
$T_L$ & ${\mathcal G}_P$\\
\hline
$ET\times_{T_L} G/T$& $\mathcal A\times_{{\mathcal G}_P}P$\\
\hline
$\Gamma(G\times _{T_R} S_{({\mathfrak g /\mathfrak t})_R})$& $\Gamma(P\times_G S)$  \\ 
\hline
\end{tabular}
\end{center}

The last line of this analogy is the most problematic. In the $G$ case
${\mathfrak g /\mathfrak t}_R$ is a finite dimensional space one can
use to construct $\mathfrak n_+$ and then get a finite dimensional complex using either
$\Lambda^*(\mathfrak n_+)$ or the spinors $S_{){\mathfrak g /\mathfrak t})_R}$.
In the $Aut(P)$ case there is no finite dimensional analog of $\mathfrak n_+$.  One can
still construct an analogous spinor bundle and it will have a space of sections that 
$Aut(P)$ acts on, but one cannot get non-trivial irreducible representations as subspace
of this space of section.  Quantum field theory indicates a method for dealing with
this problem, that of \lq\lq Second Quantization", i.e. consider that space of sections
as a space that is still to be quantized.  The space of sections will be the \lq\lq one-
particle space" and one wants to construct what can variously be thought of as the Fock
space or infinite dimensional spinor space associated to this one-particle space.

\subsection{Hilbert Spaces and Path Integrals for the Dirac Action}

A quantum field theory should associate to a manifold with boundary $(M,\partial M)$ a
Hilbert space ${\mathcal H}_{\partial M}$ which is associated to the boundary. It should
have a vacuum vector $|0>_M\in {\mathcal H}_{\partial M}$ which depends on the bounding
manifold $M$, much the way a highest weight vector in a representation of $G$ depends upon
the extra data of a complex structure on $G/T$.

We have
$$i:\partial M\rightarrow M$$
and a corresponding map
$$i^*:{\mathcal A}_M\rightarrow {\mathcal A}_{\partial M}$$
given by restriction of connection to the boundary.  Somehow we want to construct a push-forward map
$$i_!:K_{\mathcal G}({\mathcal A}_M)\rightarrow K_{\mathcal G}({\mathcal A}_{\partial M})$$
which should give us our ${\mathcal H}_{\partial M}$. We somehow have to deal with the 
fact that our ${\mathcal H}_{\partial M}$ should be a representation of 
${\mathcal G}_{\partial M}$ rather than ${\mathcal G}_M$, these groups are related by the exact sequence
$$1\rightarrow {\mathcal G}_{\partial M,1}\rightarrow {\mathcal G}_M\rightarrow {\mathcal G}_{\partial M}\rightarrow 1$$
where ${\mathcal G}_{\partial M,1}$ is the subgroup of ${\mathcal G}_M$ of gauge transformations that are the identity
on $\partial M$. The gauge group on the boundary is naturally a quotient of the gauge group on the entire space.

Recall from section \ref{spinor-geometry} that the spin representation can be defined as the 
dual space to $\Gamma(Pf^*)$ where $Pf^*$ is a line bundle over $\mathcal J (V)$, the space of
maximal isotropic complex subspaces of $V_{\mathbf C}$.  The line bundle $Pf^*$ can be explicitly
constructed on an open subset of $\mathcal J (V)$ by associating to an isotropic subspace a
skew operator and thus a vector in an exterior algebra according to the Gaussian formula
of section \ref{spinor-vacuum-vector}.

In \cite{Witten2} Witten shows that, for the case of $\Sigma$ a Riemann surface bounded by
$\partial\Sigma =S^1$,  the Hilbert space for the theory should be thought of as a space of
sections of a bundle $Pf^*$ over the space of maximal isotropic subspaces of the space of
spinor field restricted to boundary $\partial \Sigma$.  This construction is an infinite
dimensional generalization of that of section \ref{spinor-geometry} and \ref{spinor-vacuum-vector}.
The formula that associates a vector in an exterior algebra to an isotropic subspace is now
just the path integral for the Dirac action
$$\int [d\Psi] e^{\int_\Sigma \Psi \Dirac \Psi}$$
Other vectors in the exterior algebra come from evaluating the path integral with non-trivial
operator insertions
$$\int [d\Psi] e^{\int_\Sigma \Psi \Dirac \Psi}\mathcal O$$

In higher dimensions one has the same formal structure (at least for the case of $M$ even
dimensional, $\partial M$ odd dimensional): one can use the Dirac operator to polarize the
space of spinor fields restricted to $\partial M$ and interpret the fermionic path integral
as providing a construction of a bundle $Pf^*$ whose sections are the Hilbert space of the
theory, including a canonically defined vacuum vector.

\section{0+1 Dimensions: Representation Theory and Supersymmetric Quantum mechanics}

Quantum mechanics is the simplest example of a quantum field theory, one with
a zero-dimensional space and one-dimensional time.  A path integral formulation
of quantum mechanics will involve an integration over paths in some finite dimensional
manifold $M$.  A beautiful interpretation of supersymmetric quantum mechanical path integrals due
to Witten \cite{Atiyah3} interprets them as an integration map in the $S^1$-equivariant cohomology
of the loop space $LM$, where the $S^1$ action is just rotation along the loop.  Cohomological
orientability of the loop space $LM$ is equivalent to K-theory orientability of the manifold $M$.

In $S^1$-equivariant cohomology there is a localization principle that allows the calculation
of integration maps using just data from the neighborhood of the fixed point set of the $S^1$ action.
In the $LM$ case the fixed points are just the points of $M$ and the integration map reduces to
an integration over $M$, formally giving another derivation of the cohomological formula for the index.
This use of equivariant cohomology is not much related to representation theory.  The result of 
the integration map is just a constant in ${\mathbf R} [[u]]$. 

In addition, this case is rather different than that of higher dimensions.  The space-time
symmetry of this theory is dealt with by the $S^1$-equivariant cohomology, but in higher dimensions
there is no such group action and the space-time symmetry should be handled by Clifford algebra methods.

If one picks $M=G/T$, for each irreducible representation of $G$ there is a supersymmetric quantum
mechanics whose Hilbert space is just this representation. For further details about this and some
references to other work, see \cite{Woit2}.

\section{1+1 Dimensions: Loop Group Representations and Two-Dimensional Quantum Field Theories}

Two-dimensional quantum field theories involving gauge fields and fermions
 have been intensively studied over the last 20 years, partly due
to the importance of certain such theories as building blocks of conformal field theories.  Such
conformal field theories can in principle be used to construct perturbative string theories and one
might argue that they are the only part of string theory that is reasonably well-defined and
well-understood.  The Hilbert space of a two-dimensional quantum field theory depends on the
fields on the boundary of the two-dimensional base space, a set of circles.  Restricting
attention to the case of one circle, the Hilbert space will be a representation of a loop group, the
group of gauge transformations on the circle.  A crucial part of understanding the quantum field
theory is understanding how the Hilbert space decomposes into irreducible loop group representations.

There is a large literature on loop group representations, including the book \cite{PS}.  The class of irreducible loop group
representations that is both well understood and of physical significance is that of positive energy representations.  These are all
projective representations, characterized by a level $k$.

Some quantum field theories that have been studied in terms of loop group representations include:
\begin{itemize}
\item Wess-Zumino-Witten (WZW) models.  Here the fields are maps from the two dimensional base space to $G$, or equivalently gauge
transformations of a trivial bundle.  The theory is characterized by a positive integer $k$.  The Hilbert space decomposes in a similar
way to $L^2(G)$ in the Peter-Weyl theorem. It is a sum of terms of
the form
$$ \sum_{\alpha}{\overline V_\alpha}\otimes {V_\alpha}$$
where $\alpha$ is a finite set of labels of integrable $LG$ representations of level $k$.  One fascinating aspect of these
models is \lq\lq non-abelian bosonization": they are equivalent to fermionic (anti-commuting)  quantum field theories.

\item $G/H$ \lq\lq Coset" Models, or equivalently gauged Wess-Zumino-Witten models.  Here the fields are gauge transformations
of a bundle, as well as connections on the bundle.  The Hilbert space of these models is related to that of the standard WZW
model, but now one can pick out that part of the representation that is invariant under the subgroup $LH\subset LG$.

\item Supersymmetric WZW and gauged WZW models.  A supersymmetric two-dimensional quantum field theory has both anti-commuting
and commuting fields.  The Hilbert space of the theory now may be a complex.

\end{itemize}

Until recently there was little evidence in the case of these theories for the relevance of the abstract
equivariant $K$-theory point of view advocated in the earlier part of this paper.  This has changed with
the announcement \cite{Freed2} of

\begin{theorem}[Freed-Hopkins-Teleman]

For $G$ compact, simply-connected, simple, there is an equivalence of algebras
$$V_k(G)\simeq K_{G,\text {dim} G}^{k+h(G)}(G)$$
Here $V_k(G)$ is the Verlinde algebra of equivalence classes of level-$k$ positive energy representations of
$LG$ with the fusion product.  $K_{G,\text {dim} G}^{k+h(G)}(G)$ is the
$(k+h(G))$-twisted equivariant $K$-homology of $G$ (in dimension ${\text dim} G$)  with product induced from the
multiplication map $G\times G\rightarrow G$.
The $G$ action is the conjugation action, and $h(G)$ the dual Coxeter number of $G$.

\end{theorem}

While this result is expressed as a statement purely about the finite dimensional compact group $G$, it is related
to the loop group and its classifying space as follows.  Consider the trivial $G$ bundle over the circle $S^1$. In this
case the gauge group $\mathcal G$
is the loop group $LG$ and the space of connections $\mathcal A$ on the bundle has an $LG$ action.
The subgroup $\Omega G$ of based loop group elements acts freely on $\mathcal A$.  The quotient of $\mathcal A/\Omega G$ is
just $G$, the group element giving the holonomy of the connection.  The remaining $G$ action on the base point acts by
conjugation on the holonomy.  Assuming that equivariant $K-$theory can be made sense of for loop groups and that it
behaves as expected for a free action, we expect
$$K_{\mathcal G}(\mathcal A)=K_G(\mathcal A/\Omega G)=K_G(G)$$
where the $G$ action on $G$ is by conjugation.  Our discussion of
the relationship of equivariant $K$-theory and representation
theory leads us to hope for some sort of relation between the
representation ring of $LG$ and this $K_G(G)$.  The analog of the
representation ring in this context is the Verlinde algebra
$V_k(G)$ of level $k$ projective $LG$ representations, and the
Freed-Hopkins-Teleman result identifies it with $K_{G,\text {dim}
G}^{k+h(G)}(G)$.

The Freed-Hopkins-Teleman theorem fundamentally tells us that in the case of one-dimensional base space, 
the appropriate representation theory of the gauge group (that of positive-energy representations) can
be identified with the equivariant $K$-theory of the classifying space of the gauge group, although this 
requires using twisted $K$-theory since the representations are projective.  We have seen that, for a
general base space,
quantum field theory path integrals may have an interpretation as calculations in the equivariant
$K$-theory of the gauge group.  The question of whether there is an extension of the 
Free-Hopkins-Teleman result to gauge groups in higher dimensions is one of potentially great
significance for both mathematics and physics. 

Segal has given\cite{Segal-Rutgers} a map

$$V_k(G)\longrightarrow K_{G,\text {dim} G}^{k+h(G)}(G)$$ that conjecturally is the
Freed-Hopkins-Teleman isomorphism.  He constructs this map by associating to a loop group representation
$E$ the Fredholm complex 
$$E\otimes \Lambda_{\frac{1}{2}\infty}(\mathcal L\mathfrak
g^*)\stackrel{d+d^*}\longrightarrow E\otimes \Lambda_{\frac{1}{2}\infty}(\mathcal L\mathfrak g^*)$$ 
Here
$\Lambda_{\frac{1}{2}\infty}L\mathfrak g^*$ are the so-called \lq\lq semi-infinite" left-invariant
differential forms on the loop group.  He notes that this complex should be thought of as the Hilbert
space of a supersymmetric Wess-Zumino-Witten model.

This is formally similar to the Kostant complex
$$V_\lambda\otimes S^+\stackrel{\Dirac}\longrightarrow V_\lambda\otimes S^-$$
and recall that we have argued that
$$S^+\stackrel{\Dirac}\longrightarrow S^-$$
can be thought of as a $K$-homology equivariant fundamental class.  It may be possible to recast Segal's
construction in the language of Clifford algebras and spinors.  Note that the projective factor of the
dual Coxeter number contributed by the semi-infinite differential forms is the analog in the loop
group case of $\delta$ (half the sum of the positive roots) in the compact group case \cite{Freed1}.
The Kostant complex in the loop group case has been studied by Landweber \cite{Landweber2}.

While Wess-Zumino-Witten models and their supersymmetric extensions have been much studied in the physics
literature, much remains to be done to understand fully the relationship between representation theory
and these models.  The program to understand conformal field theory from representation-theoretical point of
view begun in \cite{Segal1} still remains to be developed. In particular it would be most interesting to
understand the structure of path integral calculations from this point of view.  This might provide
inspiration for how to generalize these ideas to the physical case of four dimensional space-time.

\section{Speculative Remarks About the Standard Model}

The motivating conjecture of this paper is that the quantum field theory underlying the Standard Model
can be understood in terms of the representation theory of the automorphism group of some geometric
structure. Furthermore we have argued that $K$-theory should be the appropriate abstract framework in
which to look for these representations.  The first question that arises is that of what the fundamental
geometric structure should be.  $K$-theory is well-known to have a periodicity in dimension of order 8.  
This is reflected topologically in the Bott periodicity of homotopy groups of Lie groups: for large
enough $n$, $\pi_i(Spin(n))=\pi_{i+8}(Spin(n))$. It is reflected algebraically in the periodicity of the
structure of Clifford algebra modules: there is an equivalence between irreducible modules of $C(\mathbf
R^n)$ and $C(\mathbf R^{n+8})$ (with the standard metric).  This may indicate that one's fundamental
variables can be taken to be geometrical structures on $\mathbf R^8$.

There is a long tradition of trying to use the rich structure of the Clifford algebras to
classify the particles and symmetries of fundamental physics. Perhaps a $K$-theory point
of view will allow this idea to be pursued in a more systematic way.  One aspect of the
special nature of eight dimensions is the rich geometry of $S^7$, the unit vectors in $\mathbf R^8$.
The seven sphere has no less than four distinct geometries
\begin{itemize}
\item
Real:
$$S^7=Spin(8)/Spin(7)$$
\item
Octonionic:
$$S^7=Spin(7)/G_2$$
($G_2$ is the automorphism group of the octonions $\mathbf O$).
\item
Complex:
$$S^7=Spin(6)/SU(3)=SU(4)/SU(3)$$
\item
Quaternionic:
$$S^7=Spin(5)/Sp(1)=Sp(2)/Sp(1)=Sp(2)/SU(2)$$
\end{itemize}

So by considering automorphisms of the seven-sphere one can naturally get gauge groups
$Spin(7)$, $G_2$, $SU(3)$, and $SU(2)$.  The last two are sufficient (together with overall
phase transformations) for the known Standard Model symmetries. Note that these groups are
significantly smaller than the favorite experimentally unobserved
internal symmetry groups of GUTs and string theories ($SU(5)$, $SO(10)$, $E_6$, $SO(32)$, $E_8$, etc.).

If spacetime is to be four-dimensional, we are interested in geometrical structures that
are bundles over a four-dimensional space.  Continuing to use the geometry of the seven-sphere,
perhaps one can combine one of the above geometries with the use of the fibration
\begin {equation*}
\begin{CD}
S^3 @>>>S^7 \\
@. @VVV\\
{} @. S^4
\end{CD}
\end{equation*}
so we have another $S^3=SU(2)$ internal symmetry to consider.

See \cite{Woit1} for an elaboration of some possible ideas about how this geometry
is related to the standard model.  There it is argued that
the standard model should be defined over a Euclidean signature four dimensional
space time since even the simplest free quantum field theory path integral is ill-defined
in a Minkowski signature.  If one chooses a complex structure at each point in space-time,
one picks out a $U(2)\subset SO(4)$ (perhaps better thought of as a $U(2)\subset Spin^c(4)$) 
and in \cite{Woit1} it is argued that one can
consistently think of this as an internal symmetry.  Now recall our construction of
the spin representation for $Spin(2n)$ as $\Lambda^*(\overline {\mathbf C^n})$ applied to a \lq\lq vacuum"
vector.  Under $U(2)$, the spin representation has the quantum numbers of a standard model generation
of leptons

\begin{center}
\begin{tabular}{|c|c|c|}
\hline
$\Lambda^*(\overline{\mathbf C^2})$ & $SU(2)\times U(1)$ Charges& Particles \\
\hline
\hline
$\Lambda^0(\overline{\mathbf C^2})=\mathbf 1$ & $(0,0)$& $\nu_R$\\
\hline
$\Lambda^1(\overline{\mathbf C^2})=\mathbf C^2$ & $(\frac{1}{2},-1)$& $\nu_L,\ \ e_L$\\
\hline
$\Lambda^2(\overline{\mathbf C^2})$ & $(0,-2)$& $e_R$\\
\hline
\end{tabular}
\end{center}

A generation of quarks has the same transformation properties except that one has to take
the \lq\lq vacuum" vector to transform under the $U(1)$ with charge $4/3$, which is the charge
that makes the overall average $U(1)$ charge of a generation of leptons and quarks to be zero.

The above comments are exceedingly speculative and very far from what one needs to
construct a consistent theory.  They are just
meant to indicate how the most basic geometry of spinors and Clifford algebras in low dimensions
is rich enough to encompass the standard model and seems to be naturally reflected in the
electro-weak symmetry properties of Standard Model particles.

\section{On the Current State of Particle Theory}

This article has attempted to present some fragmentary ideas relating representation theory and
quantum field theory in the hope that they may lead to new ways of thinking about quantum field
theory and particle physics and ultimately to progress in going beyond the standard model of particle
physics.  Some comments about the current state of particle theory and its problems
\cite{polemic, Friedan} may be in order
since these problems are not well known to mathematicians and their severity provides some
justification for the highly speculative nature of much of what has been presented here.

For the last eighteen years particle theory has been dominated by a single approach
to the unification of the standard model interactions and quantum gravity.  This line of thought has
hardened into a new orthodoxy that postulates an unknown fundamental supersymmetric theory involving
strings and other degrees of freedom with characteristic scale around the Planck length. By some
unknown mechanism, the vacuum state of this theory is supposed to be such that low-energy excitations
are those of a supersymmetric Grand Unified Theory (GUT) including supergravity. By another unknown
mechanism, at even lower energies the vacuum state is supposed to spontaneously break the GUT
symmetries down to those of the Standard Model and, again in some unknown way, break the
supersymmetry of the theory.

It is a striking fact that there is absolutely no evidence
whatsoever for this complex and unattractive conjectural theory.
There is not even a serious proposal for what the dynamics of the
fundamental \lq\lq M-theory" is supposed to be or any reason at
all to believe that its dynamics would produce a vacuum state with
the desired properties.  The sole argument generally given to
justify this picture of the world is that perturbative string
theories have a massless spin two mode and thus {\it{could}}
provide an explanation of gravity, {\it{if}} one ever managed to
find an underlying theory for which perturbative string theory is
the perturbation expansion.  This whole situation is  reminiscent
of what happened in particle theory during the 1960's, when
quantum field theory was largely abandoned in favor of what was a
precursor of string theory.  The discovery of asymptotic freedom
in 1973 brought an end to that version of the string enterprise
and it seems likely that history will repeat itself when sooner or
later some way will be found to understand the gravitational
degrees of freedom within quantum field theory.

While the difficulties one runs into in trying to quantize gravity in the standard way are
well-known, there is certainly nothing like a no-go theorem indicating that it is impossible to find
a quantum field theory that has a sensible short distance limit and whose effective action for the
metric degrees of freedom is dominated by the Einstein action in the low energy limit.  Since the
advent of string theory, there has been relatively little work on this problem, partly because it is
unclear what the use would be of a consistent quantum field theory of gravity that treats the
gravitational degrees of freedom in a completely independent way from the standard model degrees of
freedom. One motivation for the ideas discussed here is that they may show how to think of the
standard model gauge symmetries and the geometry of space-time within one geometrical framework.

Besides string theory, the other part of the standard orthodoxy of the last two decades has
been the concept of a supersymmetric quantum field theory.  Such theories have the huge virtue
with respect to string theory of being relatively well-defined and capable of making some predictions.
The problem is that their most characteristic predictions are in violent disagreement with
experiment.  Not a single experimentally observed particle shows any evidence of the existence
of its \lq\lq superpartner".  One can try and explain this away by claiming that an unknown
mechanism for breaking the supersymmetry of the vacuum state exists and is precisely such that
all superpartners happen to have uncalculable masses too large to have been observed.  If one 
believes this,  one is faced with the problem that the vacuum energy should then be of at least
the scale of the supersymmetry breaking.  Assuming that one's theory is also supposed to be
a theory of gravity, there seems to be no way around the prediction that the universe will be
a lot smaller than the size of a proton.

Supersymmetry has a complicated relationship with modern mathematics. The general formalism one gets by naively replacing
vector spaces by \lq\lq super vector spaces", groups by \lq\lq supergroups", etc. produces new structures but does not
obviously lead to much new insight into older mathematics. On the other hand, there certainly are crucial parts of the
mathematics discussed in this article that fit to a degree into the \lq\lq super" language. A prime example is seen in the
importance of the $\bf Z_2$ grading of Clifford algebras and its implications for $K$-theory and index theory.  In addition,
the complexes of equivariant cohomology are naturally $\bf Z_2$ graded. Whenever one works with a de Rham model of equivariant
cohomology one has something like a supersymmetry operator since the Lie derivative is a square $$\mathcal L_X=di_X
+i_Xd=(d+i_X)^2$$

Perhaps taking into account some of these other mathematical ideas can lead to new insights into which
supersymmetric quantum field theories are actually geometrically interesting and help to find new
forms of them which actually will have something to do with the real world.  

During the past twenty-five years particle physics has been a victim of its own success.  The standard
model has done an excellent job of explaining all phenomena seen up to the highest energies
that can be reached by present-day accelerators.  The advent of the LHC at CERN starting in 2007 may
change this situation but this cannot be counted on.  While historically the attempt to make progress
in theoretical physics by pursuing mathematical elegance in the absence of experimental guidance has
had few successes (general relativity being a notable exception), we may now not have any choice in
the matter.

The exploitation of symmetry principles has lead to much of the
progress in theoretical physics made during the past century.  Representation
theory is the central mathematical tool here and in various forms it
has also been crucial to much of twentieth century mathematics.
The striking lack of any underlying symmetry principle for string/M-theory is matched by the
theory's complete inability to make any predictions about nature.  This is probably
not a coincidence.

\end{document}